\documentclass[%
 nolongbibliography,
 reprint,
 amsmath,amssymb,
 aps,
 pra, 
 superscriptaddress, 
]{revtex4-2}

\usepackage{graphicx}
\usepackage{dcolumn}
\usepackage{bm}
\usepackage{lipsum}
\usepackage{braket}
\usepackage{svg}
\usepackage[breaklinks=true,colorlinks,citecolor=blue,linkcolor=blue,urlcolor=blue]{hyperref}

\newcommand{\GM}{\gamma_{\downarrow}} 
\newcommand{\GD}{\gamma_{\phi}} 
\newcommand{\trace}[1]{\operatorname{Tr} \left[ {#1} \right]}

\begin{document}

\title{
Single-atom dissipation and dephasing in Dicke and Tavis-Cummings quantum batteries}
\author{Andrea Canzio}%
\email{andrea.canzio@sns.it}%
\affiliation{NEST, Scuola Normale Superiore and Istituto Nanoscienze-CNR, I-56126 Pisa, Italy}%

\author{Vasco Cavina}%
\affiliation{NEST, Scuola Normale Superiore and Istituto Nanoscienze-CNR, I-56126 Pisa, Italy}%

\author{Marco Polini}%
\affiliation{Dipartimento di Fisica dell'Universit\`a di Pisa, Largo Bruno Pontecorvo 3, I-56127 Pisa, Italy}%
\affiliation{ICFO-Institut de Ciències Fotòniques, The Barcelona Institute of Science and Technology,
Av. Carl Friedrich Gauss 3, 08860 Castelldefels (Barcelona), Spain}%

\author{Vittorio Giovannetti}%
\affiliation{NEST, Scuola Normale Superiore and Istituto Nanoscienze-CNR, I-56126 Pisa, Italy}%

\date{\today}

\begin{abstract}
We study the influence of single-atom dissipation and dephasing noise on the performance of Dicke and Tavis-Cummings quantum batteries, where the electromagnetic field of the cavity hosting the system acts as a charger. For these models a genuine charging process can only occur in the transient regime. Indeed, unless the interaction with the environment is cut off, the asymptotic energy of the battery is solely determined by the environment and does not depend on the initial energy of the electromagnetic field.
We numerically estimate the fundamental figures of merit for the model, including the time at which the battery reaches its maximum ergotropy, the average energy, and the energy that needs to be used to switch the battery-charger interaction on and off.
Depending on the scaling of the coupling between the battery and the charger, we show that the model can still exhibit a subextensive charging time.
However, for the Dicke battery, this effect comes with a higher cost when switching the battery-charger interaction on and off.
We also show that as the number of battery constituents increases, both the Dicke and Tavis-Cummings models become asymptotically free, meaning the amount of energy that is not unitarily extractable becomes negligible.
We obtain this result numerically and demonstrate analytically that it is a consequence of the symmetry under permutation of the model.
Finally, we perform simulations for different values of the detuning, showing that the optimal regime for the Dicke battery is off-resonance, in contrast to what is observed in the Tavis-Cummings case.
\end{abstract}

\maketitle


\section{\label{sec:intro}Introduction}

Batteries have become essential for various energy-efficient applications, yet they often fall short in terms of rapid and efficient charging.
They degrade quickly and typically need replacement after a few hundred cycles.
Therefore, there is an urgent need for new energy storage technologies that go beyond traditional electrochemistry, offering fast charging capabilities while maintaining high energy density \cite{metzler2023quantum, tibben2024extendingselfdischargetimedicke}.
Starting from the seminal works of Alicki and Fannes~\cite{PhysRevE.87.042123} which introduced the idea of 
a Quantum Battery (QB)  in 2013,  several groups started considering the possibility  of exploiting quantum mechanical effects to improve the efficiency of energy storage processes \cite{Binder_2015, PhysRevLett.118.150601}.
The idea behind QBs is to harness collective and quantum effects, such as superposition and entanglement, to achieve faster energy transfer, approaching their physical limits \cite{campaioli2018quantumbatteriesreview, Bhattacharjee_2021, RevModPhys.96.031001}.

\begin{figure}[t]
    \centering
    \includegraphics[width=\linewidth]{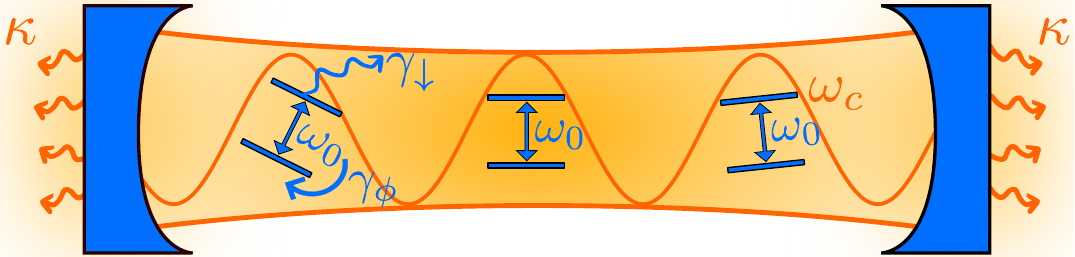}\\
    \vspace{10px}
    \includegraphics[width=\linewidth]{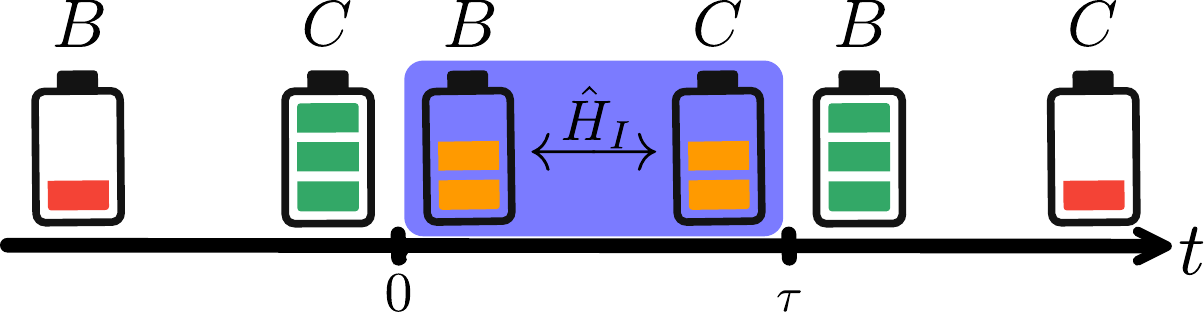}
    \caption{
        \label{fig:ndqb}Top panel: 
        Schematic of the Noisy Dicke and Tavis-Cummings models. $N$ qubits (of frequency $\omega_0$),         subject to local dephasing (rate $\gamma_\phi$) and decay (rate $\gamma_\downarrow$),
        represents the energy storage elements $B$ of the setup. They are placed in an
        imperfect optical cavity (of frequency $\omega_c$ and leakage rate $\kappa$) that acts as charger $C$. Bottom Panel: schematic representation of the charging protocol. The injection of energy  in the storing elements $B$ involves  turning on the interaction Hamiltonian $\hat{H}_I$, at $t=0$ for a duration of time $\tau$. }
\end{figure}

A model that has attracted particular attention in the last years is the Dicke Quantum Battery (DQB), where an array of $N$ two-level systems (or qubits) is charged by an optical mode through a dipole interaction, as shown in Fig.~\ref{fig:ndqb}.
Ferraro~\emph{et al.}~\cite{PhysRevLett.120.117702} showed in 2018 that such model could exhibit super-extensive charging power $P_{charg}\propto N\sqrt{N}$, a phenomenon that was qualitatively confirmed experimentally in organic-semiconductor-based Dicke quantum batteries~\cite{quach2022superabsorption}.

Since then a lot of theoretical work \cite{quach2022superabsorption, PhysRevB.98.205423, PhysRevB.99.035421, PhysRevLett.122.047702, PhysRevLett.125.236402, PhysRevB.105.115405, batteries9040197, erdman2023reinforcement, PhysRevA.109.012204, PhysRevE.107.054125, hogan2023quench} has been done on such a model.
Many tried to enhance the charged energy or reduce the charging time through parameter optimization \cite{hogan2023quench, batteries9040197, PhysRevB.105.115405} or optimal control techniques \cite{erdman2023reinforcement}.
Another crucial aspect to consider is the efficiency of converting charged energy into mechanical work. During the charging process, correlations between the charger and the battery can cause some of the energy to become locked in these correlations, making it unavailable for extraction. Significant progress has been made in understanding the fraction of energy, known as ergotropy \cite{Allahverdyan_2004}, that can be effectively harnessed \cite{PhysRevLett.122.047702, PhysRevA.109.012204}.
Further but limited work has been done in understanding whether and how noise could affect the model \cite{PhysRevB.99.035421, quach2022superabsorption}.

Here, we generalize these findings and study the feasibility of charging protocols when local noises acting on the battery and the charger are taken into account.
We study and compare the performances of two different charging interaction mechanisms, i.e.~the Dicke interaction and its rotating wave approximation, the Tavis-Cummings interaction.
After a description of the models (Sec.~\ref{sec:dicketcmodels}) we discuss possible charging protocols and the stability of the stored energy (Secs.~\ref{sec:charging} and \ref{sec:feasibility}).
Dephasing can be used as a mean of charge stabilization \cite{quach2020using, shastri2024dephasing} preventing the energy back flow from battery to charger.
When dephasing is not strong enough, another possibility is to quench off the interaction between battery and charger. In Sec.~\ref{subsec:timeandquench} we study the feasibility of such quenching mechanism by computing the associated energy cost.
In addition, by comparing the average energy and the ergotropy (Sec.~\ref{subsec:erg}), we quantify the amount of energy that is not extractable under unitaries, finding that it is negligible in the limit in which the number of constituents of the battery goes to infinity.
We thus generalize this result, known as asymptotic freedom \cite{PhysRevLett.122.047702}, to any scheme in which the battery state evolves through a permutation invariant dynamics.
In Sec.~\ref{subsec:optimiz} we conclude by discussing possible optimizations of the charging protocol, showing how to increase the charged energy and lower the charging time through detuning the frequencies of charger and battery or increasing the energy in the charger.
All results are then summarized in Sec.~\ref{sec:discussion}.


\section{\label{sec:dicketcmodels}The model(s)}

The QB model we study is sketched in the top panel of  Fig.~\ref{fig:ndqb}. It consists in an
 array of 
$N$ identical Two-Level Systems (TLSs or qubits) confined in an imperfect (lossy) optical cavity. 
The TLSs  represent the inner core $B$ of the QB
which allows for energy storage. The bosonic mode of the cavity instead represents the 
charger  $C$ that mediates the energy injection into the QB. The free Hamiltonian of the TLSs  is defined as  
$\hat{H}_B := \omega_0 [\hat{J}_z +(N /2)\hat{\mathbb{I}}]$ where $\hat{J}_{\alpha} := \sum_{i=1}^N \hat{\sigma}_{\alpha}^i / 2$ and $\hat{\sigma}_{\alpha}$ are the Pauli matrices ($\alpha=x,y,z$), $\omega_0$ being the energy gap between the two states of the TLSs ($\hbar=1$ throughout the article). The  free Hamiltonian of the charger is instead written as $\hat{H}_C := \omega_c \hat{a}^{\dagger}\hat{a}$, where $\hat{a}$ ($\hat{a}^{\dagger}$) is the bosonic annihilation (creation) operator, $\omega_c$ being the energy of one quantum (photon) of the mode.
In our study we focus on two possible types of $B$--$C$ interactions, mediated respectively by a   Dicke ({\rm D}) or Tavis-Cummings ({\rm TC}) Hamiltonian $\hat{H}_I^{({\rm X})}$ (${\rm X}=\text{D, TC}$),
\begin{align}\label{eq:DickeInt}
    \hat{H}_{I}^{(\text{D})}
    & := \frac{2g}{\sqrt{N}}\hat{J}_x \left(\hat{a} + \hat{a}^{\dagger}\right) ,
    \\ \label{eq:Tavisint}
    \hat{H}_{I}^{(\text{TC})}
    & := \frac{g}{\sqrt{N}}(\hat{J}_+\hat{a}+\hat{J}_-\hat{a}^{\dagger}) ,
\end{align}
where $\hat{J}_{\pm}:=\sum_{i=1}^N \hat{\sigma}_{\pm}^i$, $\hat{\sigma}_{\pm}^i = (\hat{\sigma}_x^i \pm i\hat{\sigma}_y^i)/2$. 
Throughout the analysis, we assume that these couplings are externally controlled via classical time-dependent pulses, allowing us to switch them on and off. This results in a global Hamiltonian of the form 
$\hat{H}^{({\rm X})}(t):=\hat{H}_B+\hat{H}_C+\lambda(t)\hat{H}_{I}^{({\rm X})}$, where 
$\lambda(t)\in[0,1]$ is a dimensionless function that represents the modulations of the  $B$--$C$ interaction. 

To account for decoherence and dissipation effects we adopt the Gorini-Kossakowski-Sudarshan-Lindblad 
(GKSL) master equation \cite{1976CMaPh..48..119L,Gorini1976,BRE02} formalism. Accordingly, we
write the dynamical evolution of the joint density matrix $\hat{\rho}$ of the $BC$ system as  follows: \begin{align}\label{eq:ODQBlma}
\begin{split}
    \dot{\hat{\rho}}(t) =&  \mathcal{L}^{({\rm X})} \hat{\rho}(t) :=
    -i\left[ \hat{H}^{({\rm X})}(t), \hat{\rho}(t) \right] + \kappa\mathcal{D}_{\hat{a}}[\hat{\rho}(t)] \\ &+ \sum_{i=1}^N \left( \gamma_{\downarrow} \mathcal{D}_{\hat{\sigma}^i_-} [\hat{\rho}(t)] + \gamma_{\phi} \mathcal{D}_{\hat{\sigma}^i_z} [\hat{\rho}(t)] \right)  ,
\end{split}
\end{align}
 where  
$\mathcal{D}_{\hat{\Theta}} [\hat{\rho}(t)] := \hat{\Theta} \hat{\rho}(t) \hat{\Theta}^{\dagger} - \frac{1}{2} \{\hat{\Theta}^{\dagger}\hat{\Theta},\hat{\rho}(t)\}$ are  Lindblad superoperators with $\{\hat{A},\hat{B}\} := \hat{A}\hat{B} + \hat{B}\hat{A}$. In the above expression
$\kappa$ is the leakage rate of the cavity mode, while $\gamma_\phi$ and $\gamma_\downarrow$ 
represent, respectively,  dephasing and dissipation rates of the TLSs. As in Refs.~\cite{PhysRevLett.118.123602,quach2022superabsorption, wang2022} in our analysis we describe these last two processes as local effects that arise at the level of single atoms, while we neglect
the presence of collective dissipation phenomena~\cite{kirtonreview}.
%
 
We conclude by highlighting  the presence of the factor $1/ \sqrt{N}$ in the coupling term of Eqs.~\eqref{eq:DickeInt} and \eqref{eq:Tavisint}.
One of the key features of the Dicke Quantum Battery, its sub-intensive charging time $\tau(N) \propto 1/\sqrt{N}$ \cite{PhysRevLett.120.117702, quach2022superabsorption} due to collective effects \cite{PhysRevB.99.205437, PhysRevResearch.2.023113, 10.3389/fphy.2022.1097564, PhysRevA.109.022210}, emerges only when using 
$g\rightarrow g\sqrt{N}$
  so that the $1/\sqrt{N}$ term in the coupling disappear.
The choice of scaling depends on the physical setup one is considering.
Taking as example the experiment of Quach~\emph{et al.}~\cite{quach2022superabsorption}, where molecules are placed in an optical cavity,  
a scaling w.r.t.~to $N$ like in Eqs. \eqref{eq:DickeInt} and \eqref{eq:Tavisint} 
describes the scenario where the size of the cavity is kept proportional to the number $N$ of molecules.
The other choice describes instead the scenario where the cavity has fixed size independent of the number $N$ of molecules.
In this work, we'll discuss both cases, highlighting their differences.


\section{\label{sec:charging} Charging protocol and figures of merit}

The charging protocol we consider is schematized in the bottom panel 
of Fig.~\ref{fig:ndqb}. The energy storing element $B$ and the charger $C$, initially prepared in an
independent configuration $\hat{\rho}(0)= \hat{\rho}_{B}(0) \otimes \hat{\rho}_{C}(0)$, are put in contact at time $t=0$ by  switching on the coupling term $\hat{H}_I^{({\rm X})}$ for some time $\tau$. The purpose of this interaction is to increase the  energy of $B$ at the expenses of $C$. 
In particular  we  focus on scenarios where 
 the former is originally discharged (i.e. prepared in the
 (zero-energy) ground state of $\hat{H}_B$ that we will denote as $\ket{\downarrow}^{\otimes N}_B$), while the latter  is in a pure state
of assigned mean energy (e.g. a coherent state $\ket{\alpha}_C$ or a Fock state of the cavity mode). 

The efficiency of the charging protocol is evaluated by considering two figures of merit: the  mean energy
 of $B$ computed at 
 the end of the charging period, $E_B(\tau) := Tr[\hat{H}_B \hat{\rho}_B(\tau)]$, 
and its associated ergotropy~\cite{Allahverdyan_2004} $\mathcal{E}(\tau)$.
Since
 $\hat{H}_B$ is a positive-semidefinite operator,  
 the mean energy precisely measures the total amount of energy loaded into $B$.
The ergotropy represents the fraction of this energy that can be extracted from $B$ through unitary transformations~\cite{Allahverdyan_2004, PhysRevE.87.042123, PhysRevLett.89.180402, Goold_2016, Vitagliano2018,  PhysRevLett.121.120602}.
It provides a {\it bona fide}  measure of the work that can be extracted from a quantum system and is computed as:
\begin{align} \label{eq:ergdef}
    \mathcal{E}(\tau) :=
    \trace{\hat{H}_B\hat{\rho}_B(\tau)} -
    \trace{\hat{H}_B\hat{\rho}_{B}^{\downarrow}(\tau)} \;,
\end{align}
where $\hat{\rho}_{B}^{\downarrow}(\tau)$  is the passive state associated with $\hat{\rho}_B(\tau)$, obtained by reordering the eigenvalues of
$\hat{\rho}_B(\tau)$ 
according to the eigenvectors 
of $\hat{H}_B$
in descending order 
(i.e., assigning higher populations to lower energy levels)~\cite{pusz1978passive}.

In a genuine charging process all the energy present in the final state of the battery should come from the charger. 
In the general setting in which $B$ and $C$ form an open system this is not always the case, since some energy could originate from the 
energy pumped when switching on (at time $0$) and off (at time $\tau$) the interaction $\hat{H}^{({\rm X})}_I$.
In addition, some energy could originate from exchanges due to the contact with the environment (see next section).
The mean switching energy cost reads
\begin{align}\label{eq:quenchcost}
    \begin{split}
        \delta E_{OFF\rightarrow ON}(0) &:=  \trace{\hat{\rho}(0)\hat{H}^{({\rm X})}_I} \;,\\
        \delta E_{ON\rightarrow OFF}(\tau) &:=  -\trace{\hat{\rho}(\tau)\hat{H}^{({\rm X})}_I}  \;.
    \end{split}
\end{align}
For the initial preparations considered here, the switch on energy $ \delta E_{OFF\rightarrow ON}(0)$ is always $0$, both for Dicke and Tavis-Cummings interactions.
On the contrary $\delta E_{ON\rightarrow OFF}(\tau)$ can give a non-trivial contribution to the energy balance.
Comparing such term with the energy transferred from $C$ to $B$ is fundamental to assess the performance of
the model, while discussing the role of the energy exchanged with the environment allows us to discriminate between two different qualitative cases: steady state charge and transient charge, which we will do in the following section.


\section{\label{sec:feasibility}Quenching dynamics: steady state vs.~transient charge}

\begin{figure}[t]
    \centering
    \includegraphics[width=\linewidth]{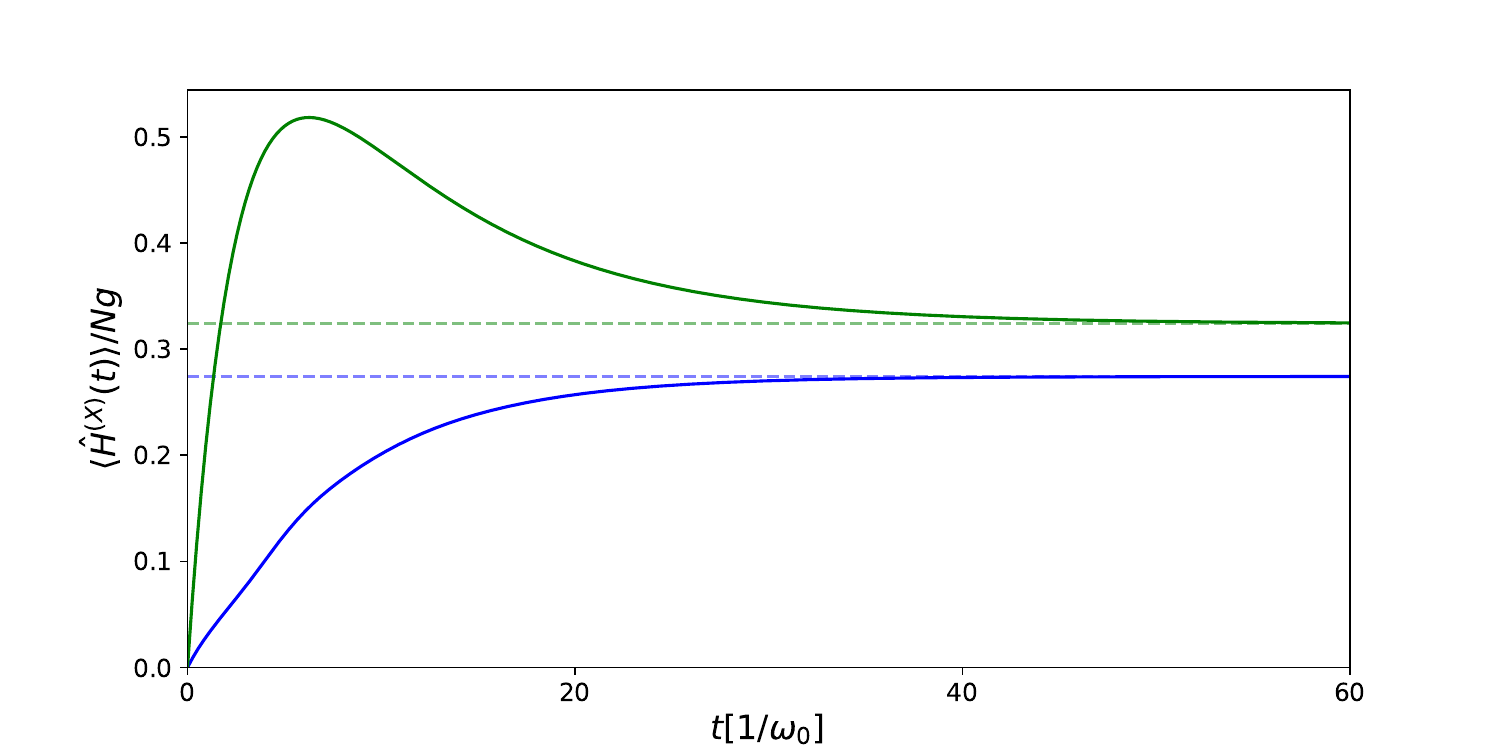}
    \caption{
        \label{fig:n4ground}
       Time evolution of the mean total energy of the system $\braket{\hat{H}^{({\rm X})}(t)}$ (battery and charger) for ${\rm X}=\text{D,TC}$ (blue solid line, green solid line) as described in Eq.~(\ref{eq:ODQBlma}) when starting from the ground state of the system, $\hat{\rho}(0)=\hat{\rho}_{gs}^{({\rm X})}$.
        The respective mean energy of the steady states are represented in dashed lines.
        Parameters: $N=4$, $\omega_0=\omega_c\equiv1$, $g=2g_c^{({\rm X})}$, $(\kappa,\gamma_\downarrow,\gamma_\phi)=(0.15,0.1,0.5)\omega_0$.
        }
\end{figure}

At difference with what previously considered for unitary energy exchanges \cite{PhysRevB.99.035421, PhysRevLett.122.047702, PhysRevA.109.012204, mazzoncini2023optimal, PhysRevB.105.115405, PhysRevE.107.054125} the presence of dissipation and decoherence in the QB model renders the 
dynamics irreversible.
In this scenario, one can talk of {\it steady state charging} when, for $\tau \rightarrow +\infty$, the battery has a non-zero amount of energy transferred \emph{from the charger}.
A necessary condition for this to happen is the existence of multiple steady states, which paves the way
to the possibility that the asymptotic charger-battery state depends on the initial preparation~\cite{albert2016geometry}, hence preserving part of the initial energy of the charger.
Examples of QBs  of this type are for instance those whose dynamical evolution is characterized by some symmetry that protect  the energy from being dissipated \cite{liu2019loss, quach2020using}.
It is well-established that both the Dicke (D) and Tavis-Cummings (TC) models exhibit a unique steady state with \emph{global} dissipators \cite{PhysRevA.95.063824, kirtonreview}.
%
We argue that the same is true in the case of local dissipators.
For the Dicke interaction this fact emerges from a numerical analysis of the dynamics generated by the associated master Eq.~(\ref{eq:ODQBlma}). The analysis reveals that, as $\tau\rightarrow +\infty$, the system reaches a unique 
steady state $\hat{\rho}_{ss}^{({\rm D})}$ which is independent from the input energy of the charger~\cite{PhysRevLett.118.123602}.
Similar considerations applies also for the TC interaction.
Many studies suggest that, irrespectively from the selected value of $g$, the steady state in the TC case is always the one with zero excitations~\cite{kirtonreview, PhysRevLett.105.043001, Larson_2017, PhysRevLett.120.183603}.
We put this observation on solid theoretical grounds presenting an explicit analytical proof of this fact (see App.~\ref{app:ssproof}).
Notice that the state with zero excitations does not necessarily coincide with the ground state of the interacting Hamiltonian $\hat{H}^{(X)}(t)$ with $\lambda(t)=1$ for either the D or TC model~\footnote{Zero temperature baths not thermalizing to the ground state may seem counter-intuitive, but it can happen since the local dissipators in Eq.~\eqref{eq:ODQBlma} do not satisfy the detailed balance conditions in respect to the total Hamiltonian $\hat{H}^{({\rm X})}(t)$ \cite{de2018reconciliation, soret2022thermodynamic}}.
As a reference see Fig.~\ref{fig:n4ground}, where we show that (for some values of $g, \kappa, \gamma_{\phi}, \gamma_{\downarrow}$) the charging of the battery occurs even when the total system is initialized in the ground state $\hat{\rho}_{gs}^{(\text{X})}$ of the total Hamiltonian.

We conclude our digression on the steady states by noting that the fact that $\hat{\rho}_{ss}^{{\rm (D,TC)}} \neq \hat{\rho}_{gs}^{{\rm (D,TC)}}$ and its consequences on the charging in Fig. \ref{fig:n4ground}, can be understood in terms of the superradiant phase transition arising in the D and TC models
when the interaction strength $g$ is above a critical value $g_c^{({\rm X})}$ \cite{HEPP1973360,kirtonreview},
\begin{align}
    \label{eq:gcrit}
    g_c^{(\text{D})} := \frac{\sqrt{\omega_0\omega_c}}{2}\; ,
    \quad
    g_c^{(\text{TC})} := \sqrt{\omega_0\omega_c} \;.
\end{align}
More specifically, when $g>g_c^{({\rm X})}$ an extensive mean number of photons is observed in the ground state of the model, $\braket{\hat{a}^{\dagger}\hat{a}}\propto \sqrt{N}$.

\begin{figure}[t]
    \centering
    \includegraphics[width=\linewidth]{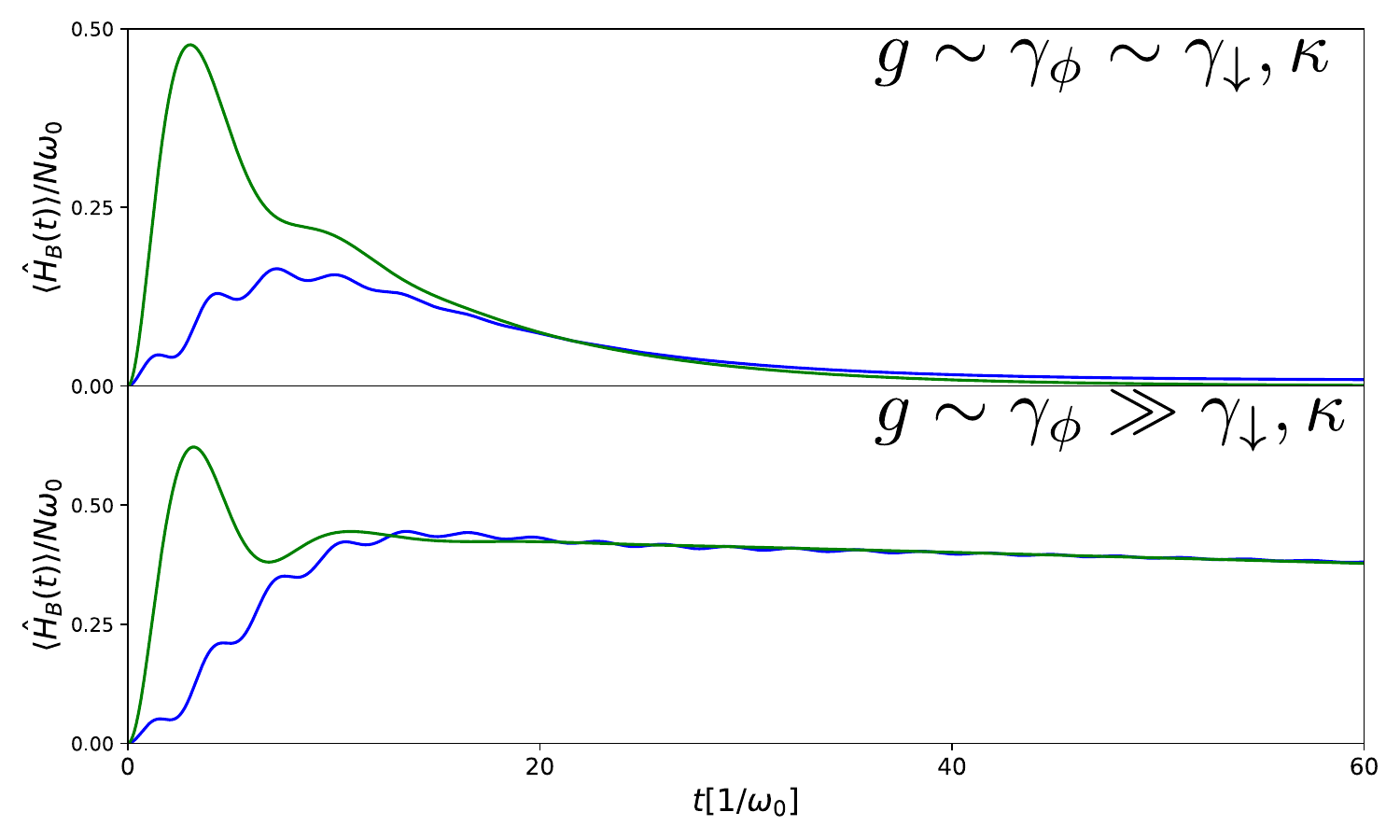}
    \caption{
        \label{fig:n4dyn}
        Time evolution of the (normalized) battery energy $\braket{\hat{H}_B(t)}$ starting from a discharged battery $\ket{\downarrow}^{\otimes N}_B$ and a charger in a coherent state $|\alpha = \sqrt{N} \rangle_C$.
        When both the coupling $g$ and dephasing rate $\gamma_\phi$ are greater then the decay channels $\kappa,\gamma_\downarrow$ the exchange of energy between the charger and the battery approximately stops within a timescale shorter than that of the decay process. Parameters and color coding as in Fig.~\ref{fig:n4ground}, with the changes $\gamma_\downarrow \rightarrow \gamma_\downarrow /20$, $\kappa\rightarrow\kappa/20$ in the bottom figure.}
\end{figure}

Based on the above discussions, we rule out steady state charging, leaving only the option of transient charging which is realized when the charging time $\tau$ is shorter than
the inverse of the  dissipation rate, i.e. $\tau \ll \gamma_{\downarrow}^{-1}$. Even if the dissipation occurs on larger time scales than the charge,
discharge naturally occurs due to a back energy flow from the battery to the charger. 
To prevent this back energy flow, we can exploit dephasing \cite{quach2022superabsorption, shastri2024dephasing}, which inhibits energy exchange between the charger and battery. Although energy will eventually dissipate into the environment due to decay, if the decay timescales are longer than those of the dephasing channel, we still observe (approximately) a stabilization of the charging process within a timescale shorter than that of the decay channel, as illustrated in Fig.~\ref{fig:n4dyn}.
Note that the dephasing rate can be artificially increased through sequential measuring  \cite{gherardini2020stabilizing}.
Another conceptually straightforward procedure consists in decoupling the battery from the charger by controlling the parameter $\lambda(t)$ in $\hat{H}_I^{({\rm X})}$, in such a case we have to ensure that the coupling energy $ \delta E_{ON\rightarrow OFF}(\tau)$ in Eq.~\eqref{eq:quenchcost} is small if compared to the energy finally stored in the battery.

\begin{figure*}[t]
    \centering
    (a)
    \includegraphics[width=0.45\linewidth]{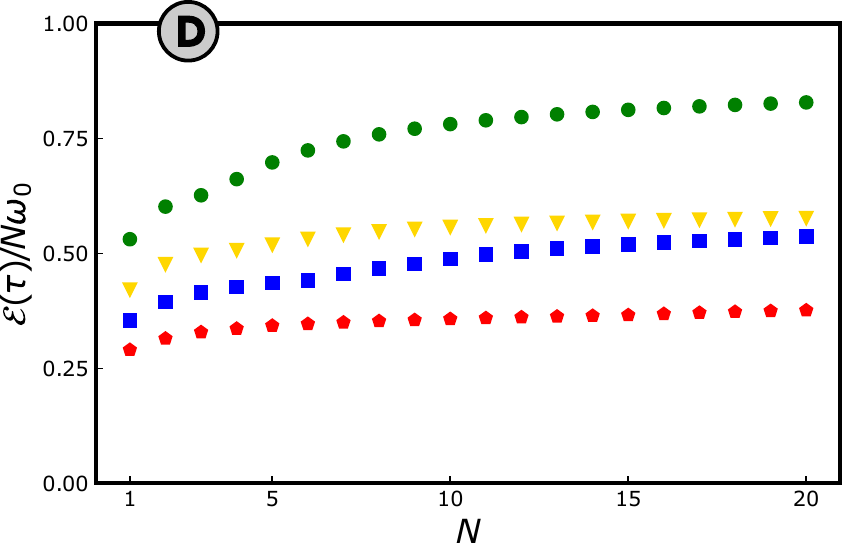}
    \quad
    (b)
    \includegraphics[width=0.45\linewidth]{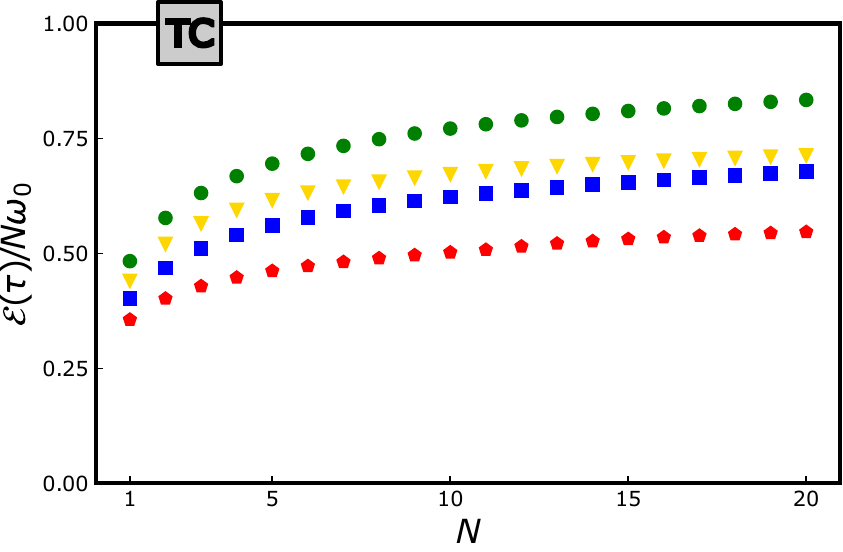}
    \\
    (c)
    \includegraphics[width=0.45\linewidth]{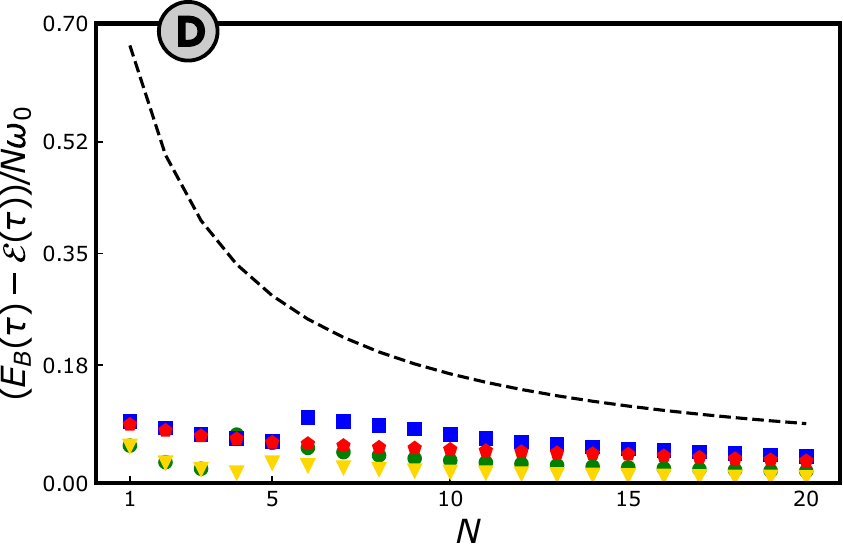}
    \quad
    (d)
    \includegraphics[width=0.45\linewidth]{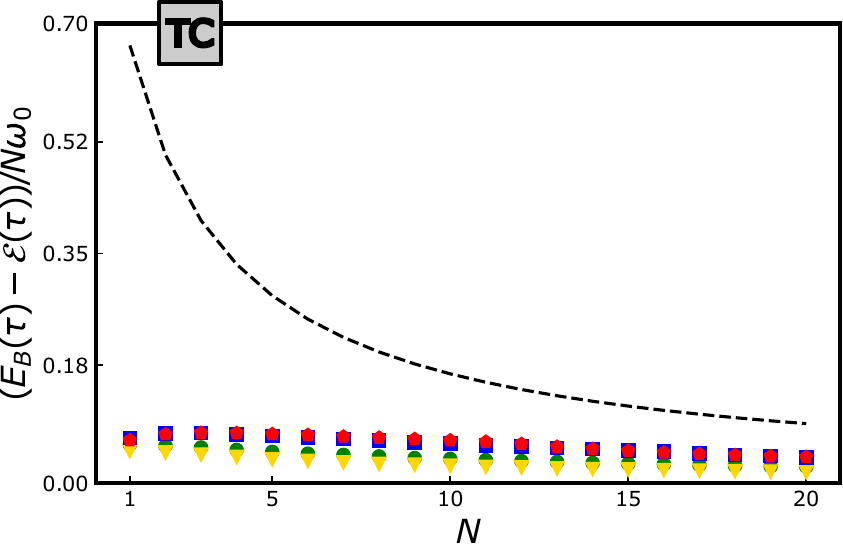}
    \caption{
        \label{fig:extr_energy}Plot of the maximum ergotropy (panels (a), (b)) and locked energy (panels (c), (d)) as a function of the number $N$ of TLSs of the battery when charging through a Dicke interaction (Eq.~\eqref{eq:DickeInt}, panels (a), (c)) or Tavis-Cummings interaction (Eq.~\eqref{eq:Tavisint}, panels (b), (d)). Here the input state of $B$ is the ground
        state  of $\hat{H}_B$ (i.e. $\ket{\downarrow}^{\otimes N}_B$)  while the charger is initialiazed in the coherent state
        $|\alpha=\sqrt{N} \rangle_C$.
        Parameters: $(\kappa,\GD,\GM)=(0,0,0)\omega_0$ (green dots); $(\kappa,\GD,\GM)=(0.15,0,0)\omega_0$ (yellow triangles); $(\kappa,\GD,\GM)=(0.15,0.1,0)\omega_0$ (blue squares); $(\kappa,\GD,\GM)=(0.15,0.1,0.1)\omega_0$ (red pentagons).
        Black line (panels (b), (d)) being the bound of Eq.~\ref{eq:ergBound}.
        Other parameters: $g =g_c^{({\rm X})}/2$ (see Eq.~\eqref{eq:gcrit}), $\omega_c=\omega_0\equiv1$.
        When the green dots are not visible, they are beneath other data points.}
\end{figure*}

\section{Results}
In this section we numerically study the interplay between quenching energy $\delta E_{ON \rightarrow OFF}(\tau)$, final energy $ E_B(\tau) $ and ergotropy $ \mathcal{E}(\tau) $ both for Dicke and Tavis-Cummings noisy batteries. 
Henceforth, the time $\tau$ will be chosen to correspond to the moment when the ergotropy of the battery is maximal.
\subsection{\label{subsec:erg}Ergotropy and asymptotic freedom}

Previous works showed that the mean energy of the battery after the charging process, $E_B(\tau)$, is extensive \cite{PhysRevLett.120.117702}, and that when charged through the Tavis-Cummings interaction the ``locked" (i.e. non-extractable) energy $E_B(\tau) - \mathcal{E}(\tau)$ is negligible in the large $N$ limit, a property also known as asymptotic freedom \cite{PhysRevLett.122.047702}.
We show that dissipation do not alter these considerations in the case of TC interactions and confirm that the results hold also for the Dicke interaction (see Fig. \ref{fig:extr_energy}).
Furthermore, we analitically demonstrate that this is a far more general property, that holds in every scenario in which the initial preparation of the battery state and the generator of the dynamics are permutation invariant.
Indeed, let us assume
\begin{equation} \label{eq:assum1}
\hat{\rho}_B(0) = \ket{\Psi} \bra{\Psi}, \, \: {\rm with} \, \: \hat{\pi} \ket{\Psi} = \ket{\Psi} ,
\end{equation}
for all permutations $\hat{\pi}$ acting on the indices of the spins composing the battery $\hat{\pi} \ket{\psi_1} \ket{\psi_2}... \ket{\psi_N} = \ket{\psi_{\pi(1)}} \ket{\psi_{\pi(2)}}... \ket{\psi_{\pi(N)}}$.
In addition, we consider a permutation invariant dynamics, that is, a generator satisfying the condition
\begin{equation} \label{eq:assum2}
\hat{\pi} \mathcal{L} [\hat{\pi}...\hat{\pi}] \hat{\pi}  = \mathcal{L}.
\end{equation}
Under the assumptions \eqref{eq:assum1} and \eqref{eq:assum2} we have
\begin{align}\label{eq:ergBound}
 E_B(\tau) - \mathcal{E}(\tau) \leq \frac{2N}{N+2}\omega_0 < 2\omega_0 ,
\end{align}
so that in the limit $N\rightarrow +\infty$, if $E_B(\tau)\propto N$, all the energy charged is always extractable via unitaries.
The bound of Eq.~\eqref{eq:ergBound} is represented in the panels~(c) and (d) of Fig.~\ref{fig:extr_energy} 
by the black dotted line.
A proof of Eq.~\eqref{eq:ergBound} can be found in App.~\ref{app:ergProof}
where we also show that the asymptotic freedom breaks if instead of considering generic unitaries as a mean of work extraction, we restrict to the much narrower class of permutation invariant unitaries.
In this case, the portion of energy that is not extractable is extensive in the large $N$ limit.

\subsection{\label{subsec:timeandquench}Charging time and quenching energy}

\begin{figure*}[!t]
    \centering
    (a)
    \includegraphics[width=0.45\linewidth]{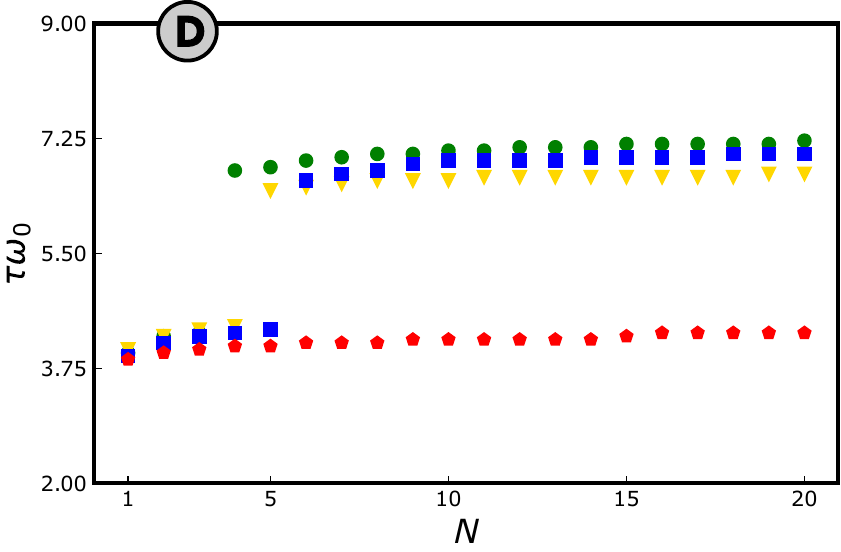}
    \quad
    (b)
    \includegraphics[width=0.45\linewidth]{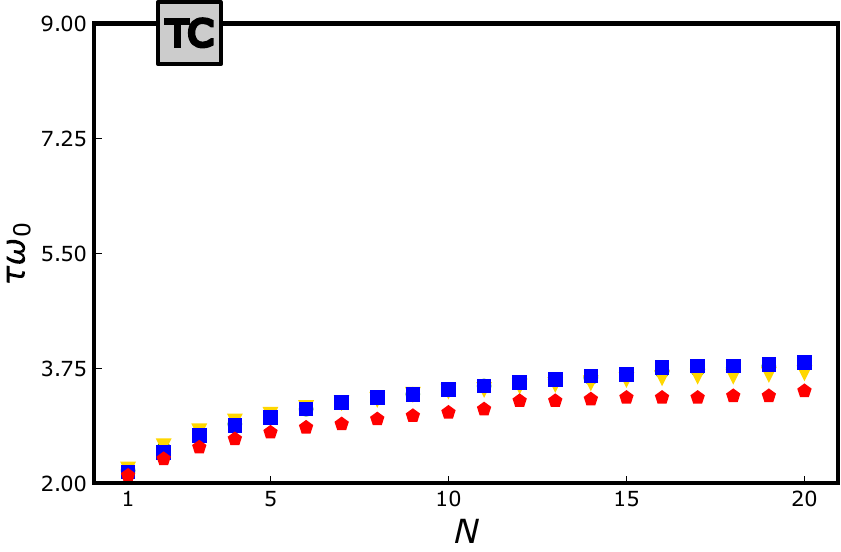}
    \\
    (c)
    \includegraphics[width=0.45\linewidth]{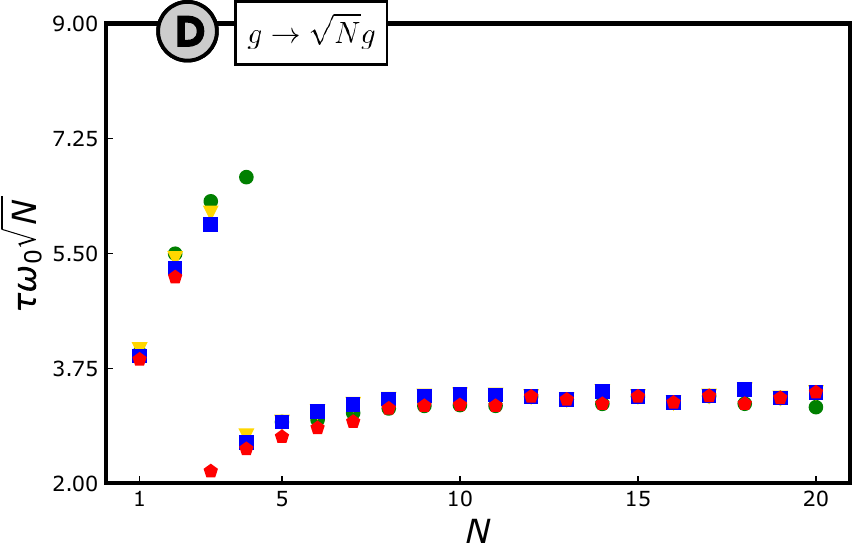}
    \quad
    (d)
    \includegraphics[width=0.45\linewidth]{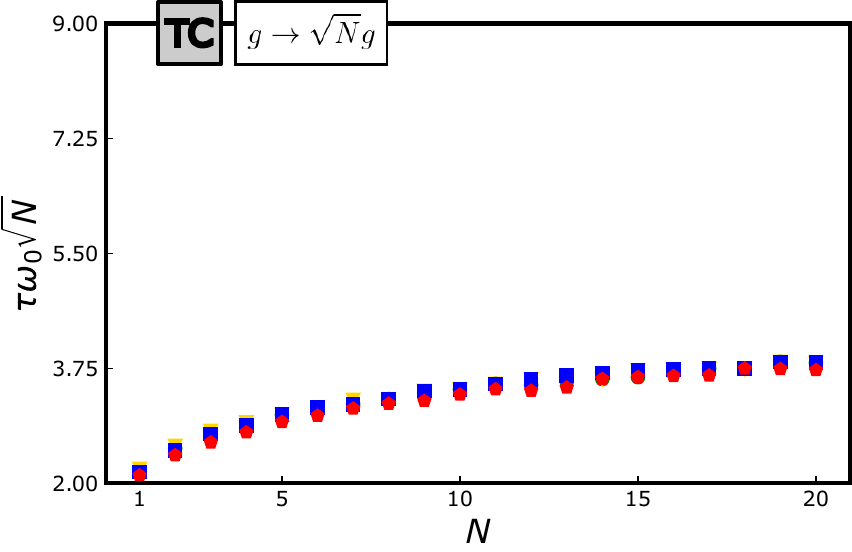}
    \caption{ Plot of the optimal charging time  as a function of the number $N$ of TLSs when charging through a Dicke interaction (Eq.~\eqref{eq:DickeInt}, panels  (a), (c)) or Tavis-Cummings interaction (Eq.~\eqref{eq:Tavisint}, panels (b), (d)).
    Panels (a), (b) has all parameters and color labeling as in Fig.~\ref{fig:extr_energy}, while panels (c), (d) has the different normalization $g\rightarrow g\sqrt{N}$ for the interaction.
    In this second scenario, the time plotted is scaled by $\sqrt{N}$ to highlight its dependence on the number of qubits $N$.
    In all the plots  the input state of $B$ is the ground state  of $\hat{H}_B$ (i.e. $\ket{\downarrow}^{\otimes N}_B$)  while the charger is initialized in the coherent state $|\alpha=\sqrt{N} \rangle_C$.
    When the green dots are not visible, they are beneath other data points.
        }
    \label{fig:times}
\end{figure*}

One of the main features of Dicke batteries is the sub-extensive charging time that is obtained under the replacement $g\rightarrow \sqrt{N} g$ in Eq.~\eqref{eq:DickeInt} \cite{PhysRevB.99.205437, PhysRevResearch.2.023113, 10.3389/fphy.2022.1097564, PhysRevA.109.022210}.
Figure~\ref{fig:times} shows the charging time for both the interactions in the case of Eqs. \eqref{eq:DickeInt} and \eqref{eq:Tavisint} (panels (a), (b)) and after rescaling $g \rightarrow \sqrt{N} g$ (panels (c), (d)).
As one can see the noises do not affect the scaling properties of the charging time $\tau$, that is constant in $N$ when charging through the interactions \eqref{eq:DickeInt}, \eqref{eq:Tavisint} (this is compatible with what found experimentally in a similar setup by Quach~\emph{et al.}~\cite{quach2022superabsorption}) and sub-extensive in the case in which $g \rightarrow \sqrt{N} g$.
%
Several interesting features can be highlighted by examining Fig. \ref{fig:times}.
First of all, in some case the charging process can receive a {\it speed up} due to the presence of the dissipators, as it is particularly evident in panel (a). 
However, this behavior does not entail a true advantage but simply is an indicator of the fact that, being the maximum ergotropy much smaller in the presence of dissipation, it is also faster to reach.
Secondly, we note a discontinuity in the charging process when the Dicke interaction is used, particularly evident - again - in panel (a).
This feature is a direct consequence of the charging protocol we designed: the charging time is chosen to ensure the maximum ergotropy, which is usually the first local maximum, but changing parameters (such as $N$) we observe that the second local maximum can become the global maximum, hence the jump in the time.
Note that while the scaling laws are the same the Tavis-Cummings interaction always outperform the Dicke interaction, having both shorter charging times and higher ergotropies.

\begin{figure*}
    \centering
    (a)
    \includegraphics[width=0.28\linewidth]{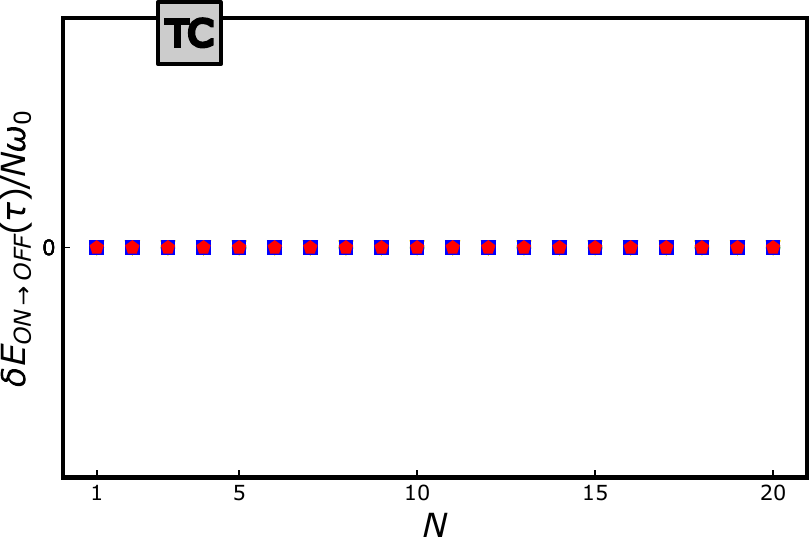}
    \quad
    (b)
    \includegraphics[width=0.28\linewidth]{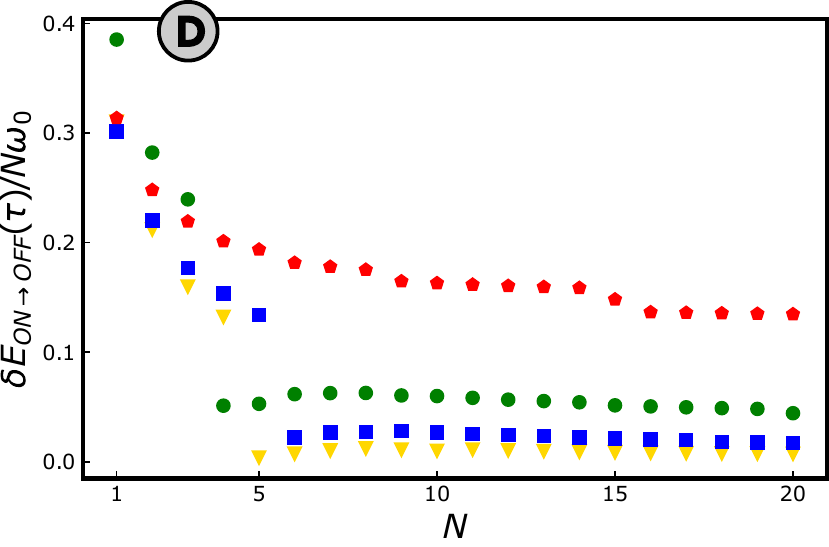}
    \quad
    (c)
    \includegraphics[width=0.28\linewidth]{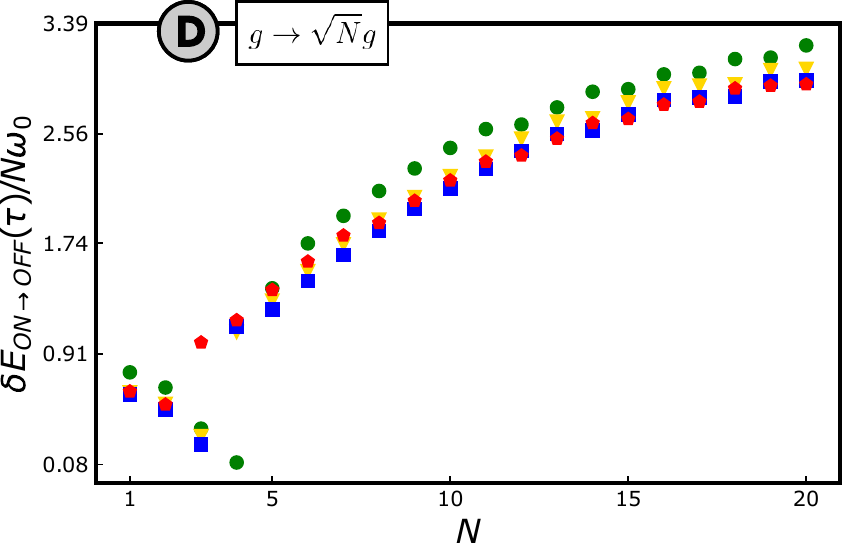}
    \caption{
        \label{fig:dEon-off}
        The dependences of the quenching energy cost $\delta E_{ON\rightarrow OFF}(\tau)$ (Eq.~\eqref{eq:quenchcost}) upon the mean number $N$ of TLSs of the battery when charging through a Dicke interaction (Eq.~\eqref{eq:DickeInt}, panels (b), (c)) or Tavis-Cummings interaction (Eq.~\eqref{eq:Tavisint}, panel (a)).
        Panels (a), (b) has all parameters and color labeling as in Fig.~\ref{fig:extr_energy}, while panel (c) has a different normalization $g\rightarrow g\sqrt{N}$ for the interaction.
        In all the plots  the input state of $B$ is the ground state of $\hat{H}_B$ (i.e. $\ket{\downarrow}^{\otimes N}_B$)  while the charger is initialized in the coherent state $|\alpha=\sqrt{N} \rangle_C$.
        }
\end{figure*}

The subextensivity of time represents an advantage of 
the case in which $g \rightarrow \sqrt{N} g$, however, in the Dicke model, the advantage in charging time is offset by the higher cost of disconnecting the charger and the battery.
We have plotted the interaction quenching cost $\delta E_{ON\rightarrow OFF}(\tau)$ in Fig. \ref{fig:dEon-off}.
When charging through the TC interaction (Fig.~\ref{fig:dEon-off} (a)) the mean energy cost for the quenching is always zero.
On the other hand, we observe a non-zero amount of energy required to turn off the Dicke interaction.
Such amount of energy is either extensive (Fig.~\ref{fig:dEon-off} (b)) in $N$ - when the interaction is normalized as in Eq.~\eqref{eq:DickeInt} - or superlinear (Fig.~\ref{fig:dEon-off} (c)) - when the interaction is rescaled as $g\rightarrow g\sqrt{N}$.
The higher energy cost in the quenching mechanism with bare coupling arises because increasing $N$ leads to a phase transition (see Sec.~\ref{sec:feasibility}), causing a significant difference between the energy eigenvectors when the interaction is active or inactive, thus requiring substantial energy to toggle the interaction.

\subsection{\label{subsec:optimiz}Optimization in respect to detuning and 
mean energy of the charger}
\begin{figure*}[!t]
    \centering
    (a)
    \includegraphics[width=0.45\linewidth]{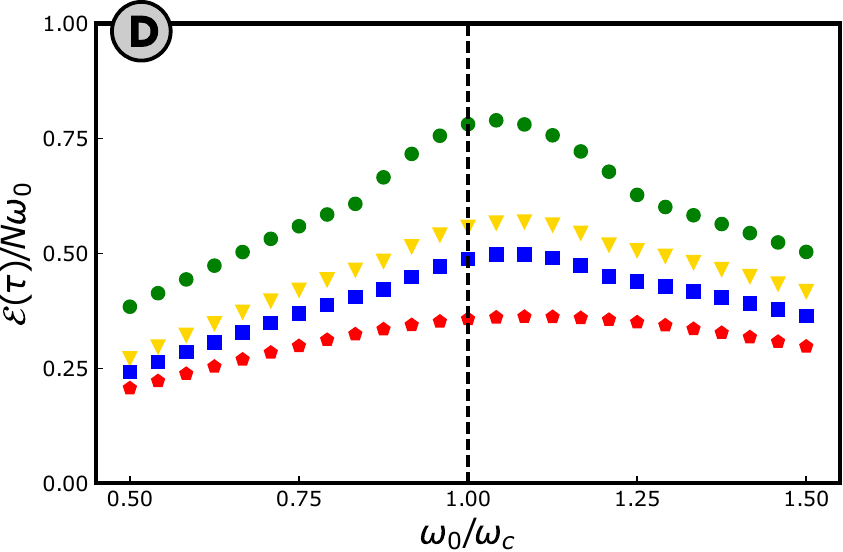}
    \quad
    (b)
    \includegraphics[width=0.45\linewidth]{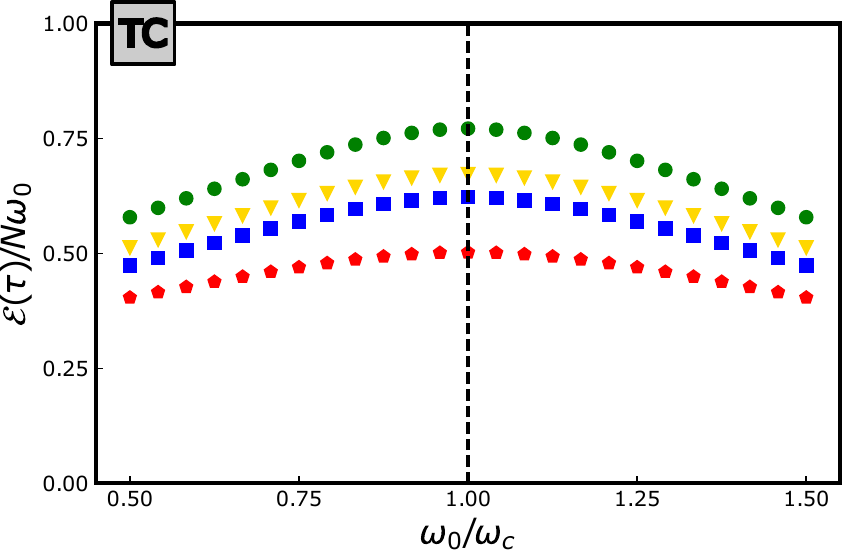}
  \\
    (c)
    \includegraphics[width=0.45\linewidth]{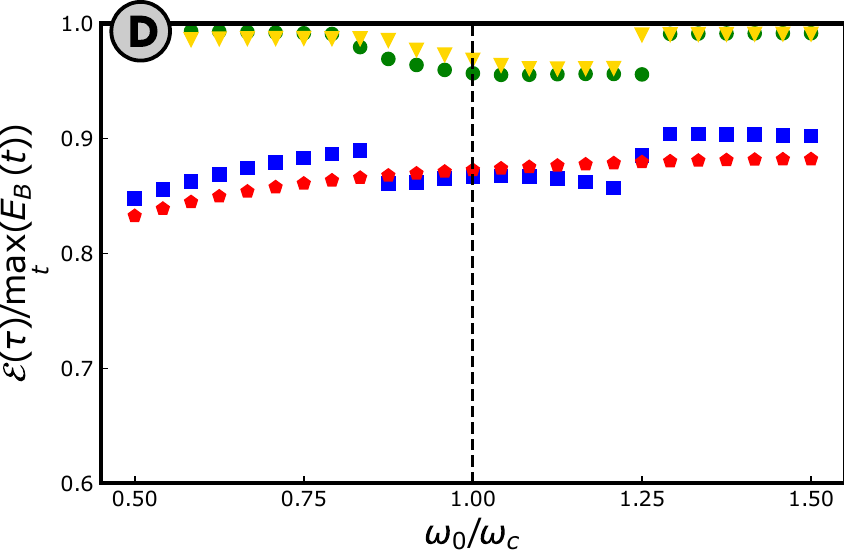}
    \quad
    (d)
    \includegraphics[width=0.45\linewidth]{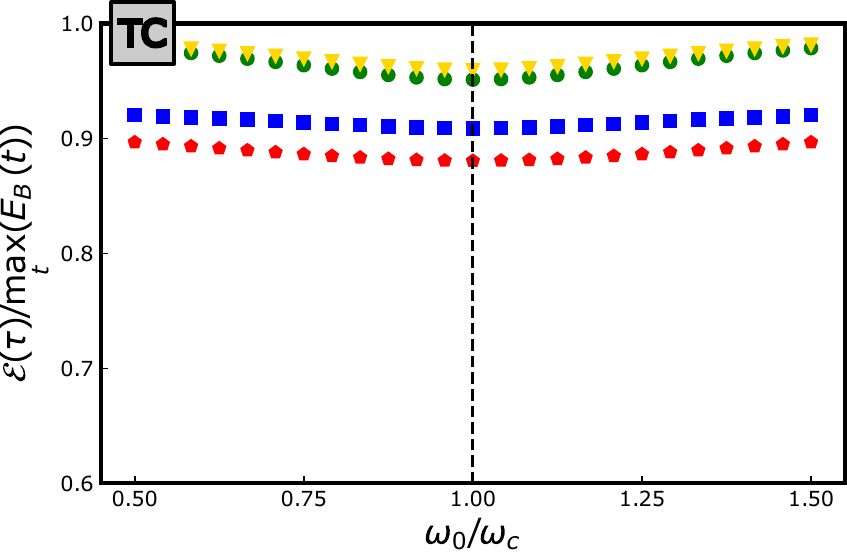}
    \caption{
        \label{fig:vsW}
        The dependences of ergotropy (panels (a), (b)) and the fraction of extractable energy (panels (c), (d)) 
        upon the battery-charger detuning $\omega_0/\omega_c$, when charging through a Dicke interaction (Eq.~\eqref{eq:DickeInt}, panels (a), (c)) or Tavis-Cummings interaction (Eq.~\eqref{eq:Tavisint}, panels (b), (d)).
        Parameters as in Fig.~\ref{fig:extr_energy} with the exception of $g$, which is half.
        $N=10$.
        In all the plots  the input state of $B$ is the ground
        state  of $\hat{H}_B$ (i.e.~$\ket{\downarrow}^{\otimes N}_B$)  while the charger is initialized in the coherent state $|\alpha=\sqrt{N} \rangle_C$.
        }
\end{figure*}

So far we studied the charging properties of a noisy battery on resonance ($\omega_0 = \omega_c$) and with the charger having as initial energy the maximum battery energy ($m\omega_c = N\omega_0$).
In the following we will observe what happens to  the performance of the battery by detuning $\omega_0$ and $\omega_c$ or varying the initial energy of the charger.
Our qualitative discussion paves to way to an optimization of the charged energy, charging time and mean power.
In Fig.~\ref{fig:vsW} we plotted the ergotropy at time $\tau$ (top panels) 
against the relative frequency between charger and battery.
We observe that when charging through the Dicke interaction 
the maximum of the ergotropy is not achieved on resonance but for $\omega_0>\omega_c$.
This implies that if we send photons at lower frequency (and, in turn, a smaller amount of energy) we can charge the battery more efficiently. 
This effect is absent for the Tavis-Cumming interaction, for which the maximum ergotropy is always at resonance.
In the bottom panels of Fig. \ref{fig:vsW} we plot the ratio of the ergotropy and the maximum energy (over time) that can be extracted from the battery.
Since we have 
\begin{equation}
    \frac{\mathcal{E}(\tau)}{\max_t E_B(t)} \leq \frac{\mathcal{E}(\tau)}{ E_B(\tau)} \leq 1 
\end{equation}
if the plotted ratio is close to one we can extract almost all the energy from the battery  by using unitary operations.
From Fig. \ref{fig:vsW} we see that the ratio decreases close to resonance both for Dicke and Tavis-Cummings interactions.
Note also that the charging time is always \emph{maximum} on resonance, so that the detuning also enhance the charging power (see appendix \ref{app:extraresults} for such plots).

\begin{figure*}
    \centering
    (a)
    \includegraphics[width=0.45\linewidth]{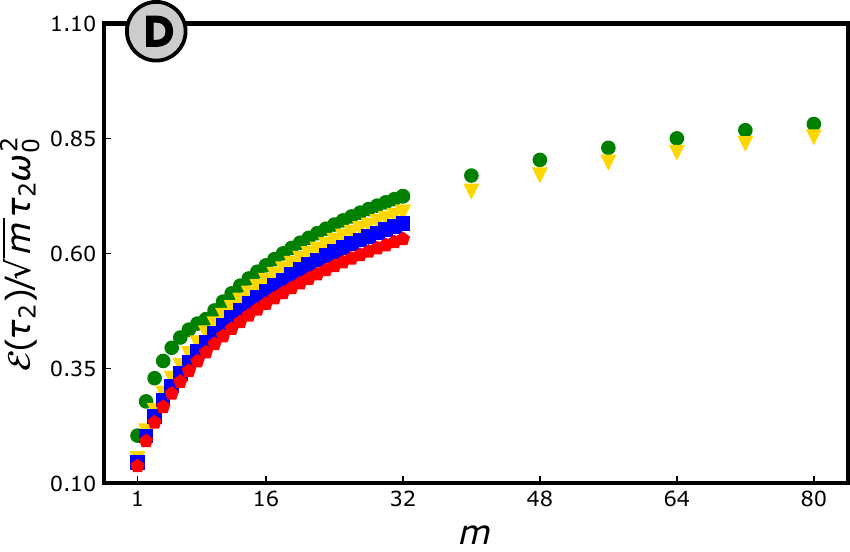}
    \quad
    (b)
    \includegraphics[width=0.45\linewidth]{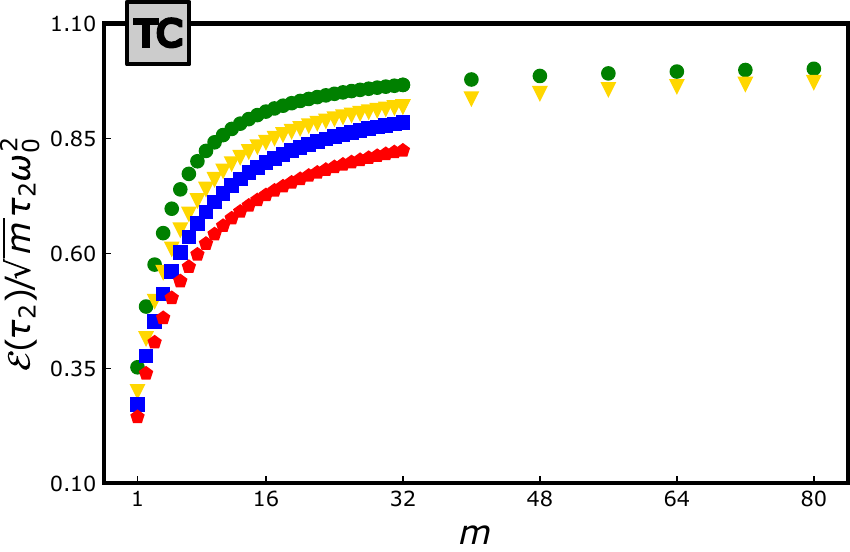}
    \caption{
        \label{fig:vsM}
        The dependences of the maximum mean power upon the charger mean energy $m\omega_c$, when charging through a Dicke interaction (Eq.~\eqref{eq:DickeInt}, panel (a)) or Tavis-Cummings interaction (Eq.~\eqref{eq:Tavisint}, panel (b)).
        Differently from the other plots, here the charging time $\tau_2$ is choosen to maximize the mean charging power, not the ergotropy.
        Parameters as in Fig.~\ref{fig:extr_energy}.
        $N=8$.
        In all the plots  the input state of $B$ is the ground
        state of $\hat{H}_B$ (i.e. $\ket{\downarrow}^{\otimes N}_B$) while the charger is initiliazed in the coherent state $|\alpha=\sqrt{m} \rangle_C$.
        }
\end{figure*}

We conclude by showing how the charging maximum mean power vary as a function of the initial energy in the charger $m\omega_c$ (see Fig. \ref{fig:vsM}).
We observe convergence to a power law $P^{max}(m) \sim P^{max}_0 \sqrt{m}$, compatible with what found by other authors \cite{hogan2023quench}.
This happens because while the maximum extractable energy is only slightly increasing by increasing $m$, the charging time reduces by a factor $1/\sqrt{m}$.
This behaviour is closely linked to the one found in \cite{PhysRevLett.120.117702} $\tau(N)\propto1/\sqrt{N}$ and shown by us in Fig.~\ref{fig:times}.
As mentioned, many of the studied quantities are approximately linear in the coupling $g$, and taking a closer look to the magnitudes in the interaction, we find that in the space of states explored we have - taking as example the Dicke interaction
\footnote{We remark that the scaling laws in Eq.~\eqref{eq:DickeIntExplained} are \emph{not} true for all states, but are true for most states explored by the dynamics with our initial conditions.},
\begin{align}\label{eq:DickeIntExplained}
    \braket{\hat{H}_I^{(\text{D})}} \simeq
    \frac{2g}{\sqrt{N}}
    \overbrace{\braket{\hat{J}_x}}^{\propto \sqrt{N}}
    \overbrace{\braket{\hat{a} + \hat{a}^{\dagger}}}^{\propto \sqrt{m}} \;.
\end{align}
Hence the result $\tau(m) \propto 1/\sqrt{m}$ is in some sense dual to $\tau(N)\propto 1/\sqrt{N}$, and both must be taken into account when considering the effective strenght of the interaction.


\section{\label{sec:discussion}Conclusion}

Many charging protocols and quantum advantages are discussed in the literature for ideal quantum batteries, but real devices are subject to dissipation and decoherence which have the potential to interfere with the charging process.
In this work we discuss open Dicke and Tavis-Cummings quantum batteries, examining the impact of local dissipation and decoherence on their performance.
We show that with different scalings of the interaction term - that reflect different physical implementations - we can charge an extensive amount of ergotropy (i.e.~extractable energy) in the battery for all the strenghts of the dissipation/decoherence considered.
The asymptotic freedom \cite{PhysRevLett.122.047702} of the batteries is observed numerically for both Dicke and Tavis-Cummings models.
To justify our findings, we prove a strong result relating the asymptotic freedom in quantum batteries to the permutation invariance of the underlying dynamical evolution.
We find that dissipation and decoherence do not affect the collective advantage in the charging time found in the closed case \cite{PhysRevB.99.205437}, and while the scaling laws are the same, the Tavis-Cummings outperforms Dicke with a lower value of the charging time.
To better understand the feasibility of the charging protocols, we quantify the energetic cost of disconnecting the coupling between the battery and the charger.
For the Tavis-Cummings interaction, this cost is negligible.
However, for the Dicke interaction, it can become the primary source of energy transfer.

In conclusion, we manipulated the parameters of the model, with the goal of improving the charged energy and lowering the charging time. By doing this, we showed, among the other results, how the best performance in the Dicke battery is reached off resonance (for $\omega_0 > \omega_c$).
Our research highlights how, despite the incredible amount of work already done on the Dicke Quantum Batteries, its behaviour it's so rich that there is still plenty to discover and look for.
In the final part of the manuscript we performed optimization only on one parameter at a time keeeping the other fixed, due to the bound set by the time required in performing the numerical simulations, specifically the worst case scenarios of $N$ large and $\GD$ or $\GM$ non-zero
\footnote{When $\GD$ or $\GM$ are not zero the $\hat{J}^2$ symmetry is broken, and the Hilbert space $\mathcal{H}_B$ of the battery goes from being of dimension $O(N)$ to $O(N^2)$, becoming much more challenging for numerical simulations.}.
Multiple parameters analysis and optimization have the potential to unveil new phenomena and may be performed for example through reinforcement learning techniques \cite{erdman2023reinforcement}.


\section*{Methods}
All the numerical simulations are performed using Python packages Qutip \cite{JOHANSSON20121760, JOHANSSON20131234} and Piqs \cite{piqsmain}.
The dimension of the Hilbert space of the bosonic mode was taken to be $\dim(\mathcal{H}_C) = 4m+1$, where $m$ is the initial mean number of photon of the charger.
The codes, the raw data generated by the code, all the plots shown and more will be available at \href{https://github.com/TheQuantumJoker/ndqb}{https://github.com/TheQuantumJoker/ndqb} and are already available upon request.

\begin{acknowledgments}
We gratefully acknowledge the computational resources of the Center for High Performance Computing (CHPC) at SNS.
M.~P.~was supported by the National Quantum Science and Technology Institute (NQSTI) through the project
“Quantum many-body batteries: circuit QED architectures and NV centers in diamond”, Grant No.~J13C22000680006. 
We thank Gian Marcello Andolina and Antonio D'Abbruzzo for useful discussions.
\end{acknowledgments}


\bibliographystyle{apsrev4-2}
\bibliography{bibliography} 

\onecolumngrid
\pagebreak
\appendix
\twocolumngrid

\section{\label{app:ssproof}Steady state of the TC model with local noises}
\label{app:TCss}

The steady states of the TC model described by Eq.~\eqref{eq:ODQBlma} with ${\rm X}={\rm TC}$ can be obtained with a standard Lyapunov analysis \cite{fasano2006analytical}.
For the rest of this section, we will denote the dynamical map generated by Eq.~\eqref{eq:ODQBlma} as $\Lambda^{({\rm TC})}(t) : = e^{t \mathcal{L}^{({\rm TC})}}$ and study its restriction to the space of density matrices $\mathcal{B}$.
By definition of steady state $\hat{\rho}_{ss}^{({\rm TC})}$ we have 
\begin{equation} \label{eq:fixed}
\Lambda^{({\rm TC})}(t) \hat{\rho}^{({\rm TC})}_{ss} = \hat{\rho}^{({\rm TC})}_{ss},
\end{equation}
for all choices of $t$. There is a unique state satisfying the condition above, namely the state without bosonic or spin excitations
\begin{equation} \label{eq:noext}
    \hat{\rho}_{ss}^{({\rm TC})} =\ket{\text{\O}}_C \bra{\text{\O}} \otimes \left( \ket{\downarrow}_B\bra{\downarrow}
    \right)^{\otimes N}\;,
\end{equation}
with $|\text{\O}\rangle_C$ the vacuum state of the cavity mode. 
We first prove that the state in Eq.~\eqref{eq:noext} is a fixed point of the dynamical evolution. Then, we will show that it is actually the only one.
In both the stages of the proof, it will be convenient to make use of the operator associated to the total number of excitations
\begin{equation}
\hat{N}_{ex} = \hat{a}^{\dag} \hat{a} +\hat{L}_z + \frac{N}{2},
\end{equation}
that is a well-known invariant of the dynamics generated by the Tavis-Cummings Hamiltonian
\begin{equation} \label{eq:preserv}
[\hat{H}^{({\rm TC})}, \hat{N}_{ex}] = 0.
\end{equation}
Since $\hat{\rho}_{ss}^{({\rm TC})}$ in Eq.~\eqref{eq:noext} is the sole eigenvector of $\hat{N}$ with eigenvalue $0$, by virtue of Eq.~\eqref{eq:preserv} such state is also invariant under the dynamics generated by $\hat{H}^{({\rm TC})}$.
Focusing on the dissipative part of the evolution, one can easily verify that 
\begin{equation} \label{eq:dissipy}
\mathcal{D}[\hat{\sigma}^i_z]  \hat{\rho}_{ss}^{({\rm TC})}=\mathcal{D}[\hat{\sigma}^i_-] \hat{\rho}_{ss}^{({\rm TC})} =
\mathcal{D}[\hat{a}] \hat{\rho}_{ss}^{({\rm TC})} = 0\;. 
\end{equation}
%
%
Since $\hat{\rho}_{ss}^{({\rm TC})}$ nullifies under both the action of the dissipators (see Eq.~\eqref{eq:dissipy}) and the generator of the unitary evolution (see below Eq.~\eqref{eq:preserv}) we conclude that $ \mathcal{L}^{({\rm TC})} \hat{\rho}_{ss}^{({\rm TC})} =0$ from which Eq.~\eqref{eq:fixed} follows.

To prove the unicity, we resort on the definition of {\it Lyapunov function}.
If a function $f$ of a generic state $\hat{\rho}$ is such that
\begin{equation} \label{eq:Lyap}
    f(\hat{\rho}) \geq 0, \quad  f \big( \mathcal{L}^{({\rm TC})}\hat{\rho}\big) \leq 0 \quad \forall \hat{\rho} \in \mathcal{B}, 
\end{equation}
 where $f(\hat{\rho}) = 0$, $ f\big(\mathcal{L}^{({\rm TC})}\hat{\rho}\big) = 0$ only for $\hat{\rho} = \hat{\rho}_{ss}^{({\rm TC})}$, then $\hat{\rho}_{ss}^{({\rm TC})}$ is a unique fixed point for $\Lambda^{({\rm TC})}$.
Hence, to conclude, we have to show that a function $f$ with the properties \eqref{eq:Lyap} exists. 
Our candidate is the average number of excitations
\begin{equation}
  n_{ex}(\hat{\rho}) : = Tr [ \hat{N}_{ex} \hat{\rho}],
\end{equation}
that trivially satisfies $  n_{ex}(\hat{\rho})  > 0$ for all the states, with the exception of $\hat{\rho}_{ss}^{({\rm TC})}$ for which we have $ n_{ex}(\hat{\rho}_{ss}^{({\rm TC})}) = 0$.
It remains to prove that 
\begin{equation}
  Tr [ \hat{N}_{ex}  \mathcal{L}^{({\rm TC})} \hat{\rho}] < 0 \, \; \forall \hat{\rho} \in \mathcal{B} \, \, {\rm s.t.} \, \, \hat{\rho} \neq  \hat{\rho}_{ss}^{({\rm TC})}.
\end{equation}
Due to Eq.~\eqref{eq:preserv}, for the Hamiltonian part of $\mathcal{L}^{({\rm TC})}$ we have
\begin{equation}
   -i Tr [ \hat{N}_{ex}  [\hat{H}^{({\rm TC})}, \hat{\rho} ] ] = 
     -i Tr [ [\hat{N}_{ex} , \hat{H}^{({\rm TC})}] \hat{\rho}  ] =0.
\end{equation}
Let us compute the action of the dissipators, starting with $\mathcal{D}[\hat{a}]$, we have
\begin{equation} \label{eq:calc1}
     Tr [ \hat{N}_{ex}  \mathcal{D}[\hat{a}] \hat{\rho}] =
        Tr [ \mathcal{D}^{\dagger}[\hat{a}]\hat{N}_{ex}   \hat{\rho}]
\end{equation}
where $\mathcal{D}^{\dagger}[\hat{\Theta}] : = \hat{\Theta}^{\dag} ... \hat{\Theta} - \frac{1}{2} \{\hat{\Theta}^{\dagger}\hat{\Theta}, ... \}$ is the adjoint superoperator of $\mathcal{D}[\hat{\Theta}]$. We compute its effect on the part of $\hat{N}_{ex}$ that acts on the cavity mode, i.e. $\hat{a}^{\dag} \hat{a} $:
\begin{align}
      \mathcal{D}^{\dagger}[\hat{a}]\hat{a}^{\dag} \hat{a} =  \hat{a}^{\dag}  
 \hat{a}^{\dag} \hat{a} \hat{a} - \frac{1}{2} \{\hat{a}^{\dagger}\hat{a}, \hat{a}^{\dag} \hat{a} \} = - \hat{a}^{\dag} \hat{a}.
\end{align}
After replacing this in Eq.~\eqref{eq:calc1} we have
\begin{equation} \label{eq:calc2}
     Tr [ \hat{N}_{ex}  \mathcal{D}[\hat{a}] \hat{\rho}] = - Tr [  \hat{\rho} \hat{a}^{\dag}\hat{a} ] \leq 0
\end{equation}
where the equality holds only for states without photonic excitations.
The spin damping yields similar results
\begin{align} \notag
 \mathcal{D}^{\dagger}[\hat{\sigma}_-^i] \hat{L}_z  &=
  \sigma_+^i   \hat{L}_z \sigma_-^i- \frac{1}{2}\{ \hat{\sigma}_+^i \hat{\sigma}_-^i , \hat{L}_z\} \\  \notag
  & =  \frac{1}{2} \hat{\sigma}_+^i   \hat{\sigma}^i_z \hat{\sigma}_-^i- \frac{1}{4}\{ \hat{\sigma}_+^i \hat{\sigma}_-^i , \hat{\sigma}_z^i\}  \\ &=  
  \frac{1}{2} \hat{\sigma}_+^i  \hat{\sigma}_-^i \hat{\sigma}^i_z -  \frac{1}{2} \hat{\sigma}_+^i \hat{\sigma}_-^i - \frac{1}{2} \hat{\sigma}_+^i \hat{\sigma}_-^i  \notag \\
  &=  
  -\frac{1}{2} \hat{\sigma}_{+}^i \hat{\sigma}_{-}^i = -\frac{1}{2} \big( \hat{\sigma}_z^i + 1 \big)
  \end{align}
where we used that $ \mathcal{D}^{\dagger}[\hat{\sigma}_-^i] \hat{\sigma}_{z}^j =0 $ for $ j \neq i $ and the commutation relations for the ladder operators $[\hat{\sigma}_z, \hat{\sigma}_{\pm}^i] = \pm 2 \hat{\sigma}_{\pm} $. 
Summing over all the local dissipators we have
\begin{align}
   \sum_{i=0}^N  \mathcal{D}^{\dagger}[\hat{\sigma}_-^i] \hat{L}_z = - \frac{1}{2} \sum_{i=0}^N  \big( \hat{\sigma}_-^i + 1 \big) = -\hat{L}_z -\frac{N}{2}
\end{align}
from which we conclude that
\begin{equation} \label{eq:calc3}
    Tr[ \hat{N}_{ex} \sum_{i=1}^N \mathcal{D}[\hat{\sigma}_i] \hat{\rho}] =  - Tr[  \hat{\rho} \hat{L}_z ] - \frac{N}{2} \leq 0 ,
\end{equation}
where the equality holds only for states without any spin excitation.
After noting that the contribution of the dephasing nullifies, that is
\begin{equation}
    \mathcal{D}^{\dag}[\hat{\sigma}_z^i] \hat{N}_{ex} = 0 
\end{equation}
we can put together Eqs. \eqref{eq:calc2} and \eqref{eq:calc3} and obtain 
\begin{equation}
    Tr [ \hat{N}_{ex}  \mathcal{L}^{({\rm TC})} \hat{\rho}] = - \kappa Tr[\hat{a}^{\dag} \hat{a} \hat{\rho}] - \gamma_{\downarrow} 
     Tr[\big( \hat{L}_z + \frac{N}{2}\big)] \leq 0
\end{equation}
where the equality holds only for states without spin or bosonic excitations, i.e. only for $\hat{\rho}_{ss}^{({\rm TC})}$ in Eq.~\eqref{eq:noext}.
This proves that $ n_{ex}(\hat{\rho})$ is the Lyapunov function we were looking for, consequently  $ \hat{\rho}_{ss}^{({\rm TC})} $ is the unique steady state of the Tavis-Cummings model with local dissipators.


\section{\label{app:ergProof}Proof of Eq.~\eqref{eq:ergBound}}

The passive counterpart of any battery state $\hat{\rho}_B$, appearing in Eq.~\eqref{eq:ergdef}, is the state which has the lowest mean energy among those with the same spectrum of~$\hat{\rho}_B$.
If we introduce the spectral decompositions of a given state $\hat{\rho}_B=\sum_i \eta_i |i\rangle\langle i|$ and of its Hamiltonian
$\hat{H}_B= \sum_i \epsilon_i |\epsilon_i\rangle\langle\epsilon_i|$ 
we can write \cite{pusz1978passive}		
	\begin{eqnarray}  \label{eq:passappend}
	\hat{\rho}^{\downarrow}_B := \sum_{i} \eta^{\downarrow}_i  |  \epsilon^{\uparrow}_i \rangle\langle \epsilon^{\uparrow}_i |
\;,
	\end{eqnarray} 
	where 
	 $\eta^{\downarrow}_{\rho}:=\{ \eta^{\downarrow}_1, \eta^{\downarrow}_2, \cdots\}$ is a rearrangement of the 
	 spectrum $\eta_{\rho}:=\{ \eta_1, \eta_2, \cdots\}$
	 of $\hat{\rho}_B$ where the various terms are organized in decreasing order (i.e. $\eta^{\downarrow}_{i} \geq \eta^{\downarrow}_i$), and $\{ |  \epsilon^{\uparrow}_i \rangle\}_i$ are instead the eigenvectors of the system Hamiltonian organized in 
	 increasing order of their associated eigenvalues (i.e. $\epsilon^{\uparrow}_i \leq \epsilon^{\uparrow}_{i+1}$). 

Using the characterization of the passive state provided by Eq.~\eqref{eq:passappend}
we can derive the straightforward bound
\begin{align} \label{eq:dbound}
    Tr[\hat{H}_B \hat{\rho}_{B}^{\downarrow}] \leq \frac{1}{d}\sum_{i=0}^{d-1}\epsilon_i^{\uparrow} \;.
\end{align}
where $d$ is the dimension of the Hilbert space of the system, in our case $d= 2^N$.
However, if we make the assumption that both the system evolution and the initial state preserve permutational symmetry, then the states at all times live in the subspace $\mathcal{S}$ generated by the Dicke Basis \cite{piqsmain, cavina2024symmetry}
\begin{align}
    dim \mathcal{S} =
    \begin{cases}
        \frac{(N+2)^2}{4} & \mbox{if $N$ even},\\
        \frac{(N+2)^2}{4} - \frac{1}{4} & \mbox{if $N$ odd}.
    \end{cases}
\end{align}

Assuming $N$ even (the proof proceeds with similar steps for $N$ odd)  and using Eq.~\eqref{eq:dbound} we have that the energy of the passive state in an ensemble of $N$ two-level systems preserving permutational symmetry satisfies the following inequality,
\begin{align} \label{eq:passlim}
    Tr[\hat{H}_B \hat{\rho}_{B}^{\downarrow}] \leq \frac{4}{(N+2)^2}\sum_{i=0}^{(N+2)^2/4-1}\epsilon_i^{\uparrow} \;.
\end{align}
We know that the spectrum of the Hamiltonian is $\bar{\epsilon}_k = k\omega_0$ where each eigenvalue has multiplicity $\binom{N}{k}$, so that to cover a subspace of dimension $\dim \mathcal{S}$ we only need the first three eigenvalues, since $\binom{N}{0} + \binom{N}{1} + \binom{N}{2} \geq \frac{(N+2)^2}{4}$. Thus we have
\begin{align} \notag
         Tr[\hat{H}_B \hat{\rho}_{B}^{\downarrow}]  \leq&
    \frac{4}{(N+2)^2} \left[ 0 * \binom{N}{0} + \omega_0 * \binom{N}{1} \right. \\ 
    +&  \left. 2\omega_0 * \left( \frac{(N+2)^2}{4} - \binom{N}{0} - \binom{N}{1} \right) \right]  \label{eq:bound}
   \notag \\ 
    \leq& \frac{2N}{N+2}\omega_0 < 2\omega_0 \;,
\end{align}
which is the desired result.
Note that to obtain this result it is fundamental to assume that we can use all the unitaries in the energy extraction process.
If we assume that only unitaries preserving the permutation invariance are allowed, we cannot go outside from the Dicke basis and the multiplicity of the eigenvalues change radically.
The sum of the energies appearing in Eq.~\eqref{eq:dbound} becomes equal to
\begin{align} \notag
  & \omega_0 \sum_{l=0}^{N/2} \sum_{m=-l}^l \omega_0 \big(m +N/2 \big)  
    \\   \notag 
 = &   \sum_{l=0}^{N/2} \sum_{m=-l}^l \frac{\omega_0 N}{2}   =
     \sum_{l=0}^{N/2}  \frac{\omega_0 N}{2} (2l+1) 
       \\  =   & \sum_{l=1}^{N+1} \frac{\omega_0 N}{2} l
       =\omega_0 \frac{N (N+1)(N+2)}{4}  .
\end{align}
We conclude that the bound \eqref{eq:dbound} restitues
\begin{equation}
       Tr[\hat{H}_B \hat{\rho}_{B}^{\downarrow}] \leq \frac{1}{\dim \mathcal{S}}\sum_{i=0}^{d-1}\epsilon_i = \omega_0 \frac{N (N+1)}{(N+2)}. 
\end{equation}
The bound above for large $N$ is linear in $N$, this means that if we can only use permutation invariant unitaries it is not ensured that an extensive amount of energy is extractable.

\section{\label{app:extraresults}Fock charging and minor results}

In the main body of the article we discusses about the extractable energy, quenching mechanism, charging time of the considered charging protocol and their optimization.
All the results shown were evaluated using a charger initially in a coherent state.
The aim of this appendix is to list all the differences found in the results when the charger is initially in a Fock state and show some other minor results not discussed in the main body.

\subsection{Similarities and differences when charging through a Fock state}
\begin{figure*}
    \centering
    (a)\includegraphics[width=0.205\linewidth]{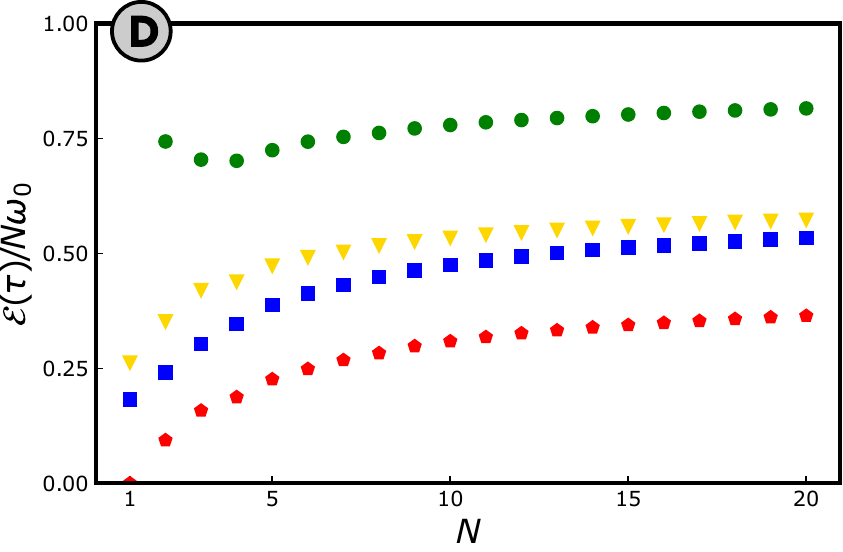}
    \quad
    (b)\includegraphics[width=0.205\linewidth]{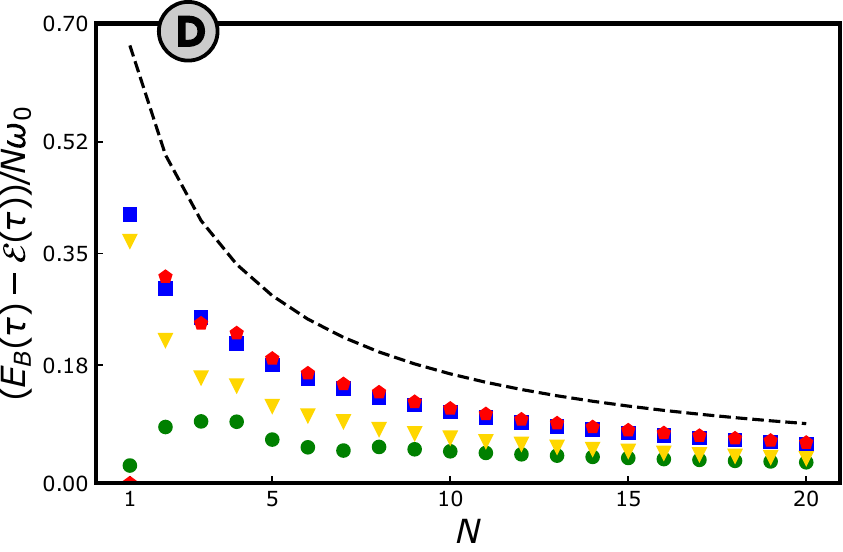}
    \quad
    (c)\includegraphics[width=0.205\linewidth]{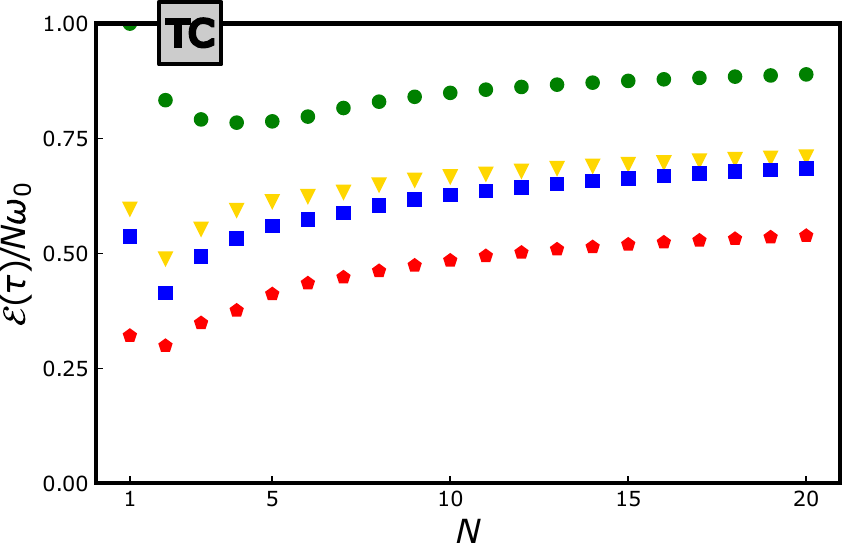}
    \quad
    (d)\includegraphics[width=0.205\linewidth]{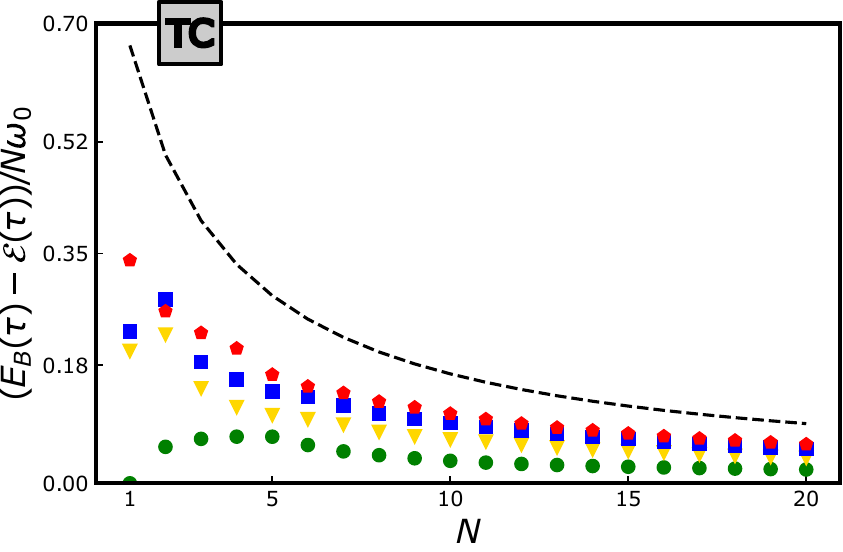}
    \\
    (e)\includegraphics[width=0.205\linewidth]{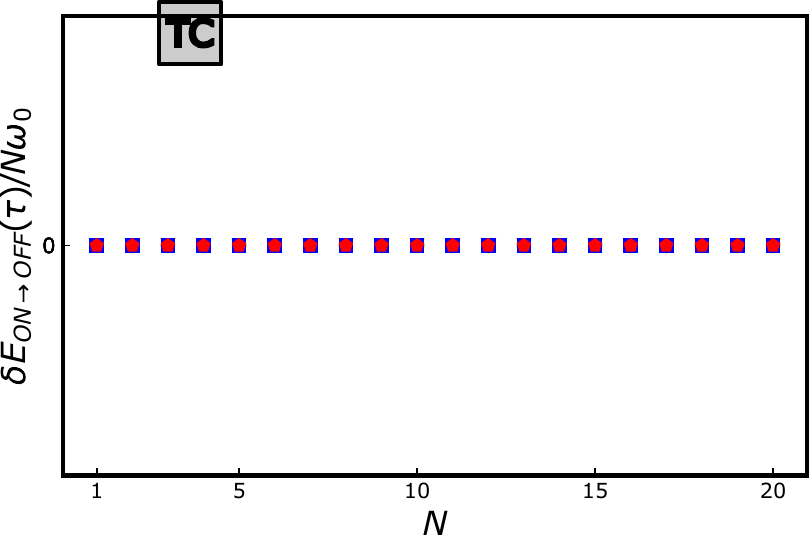}
    \quad
    (f)\includegraphics[width=0.205\linewidth]{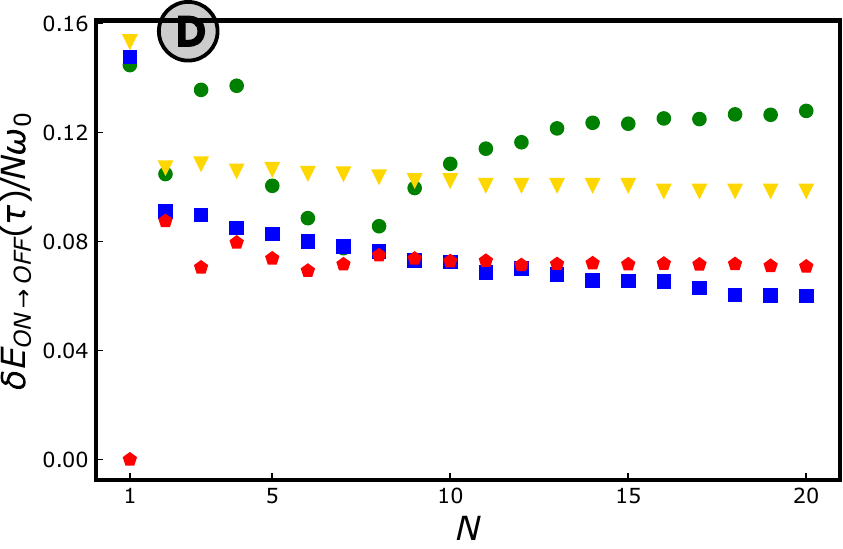}
    \quad
    (g)\includegraphics[width=0.205\linewidth]{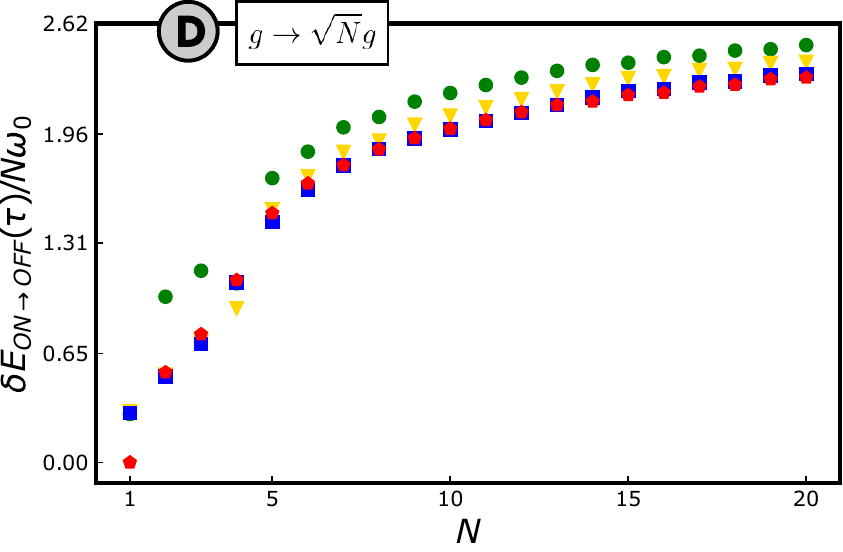}
    \\
    (h)\includegraphics[width=0.205\linewidth]{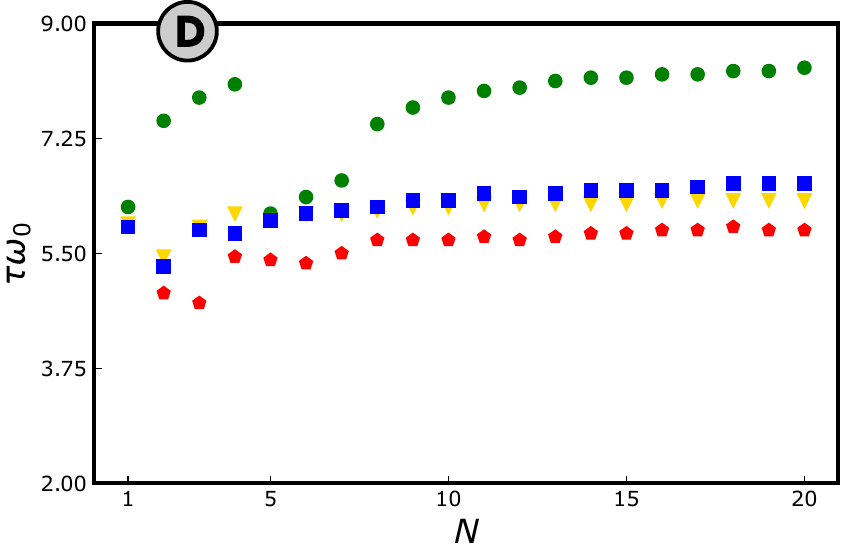}
    \quad
    (i)\includegraphics[width=0.205\linewidth]{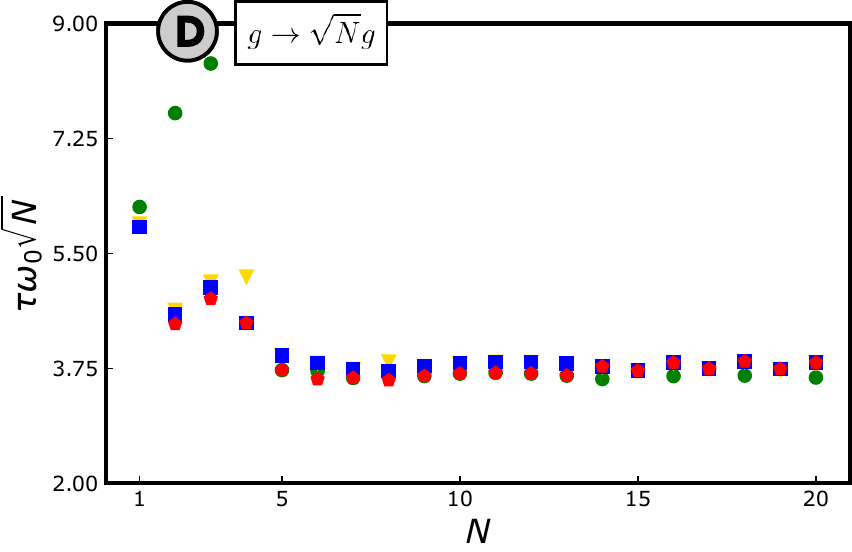}
    \quad
    (j)\includegraphics[width=0.205\linewidth]{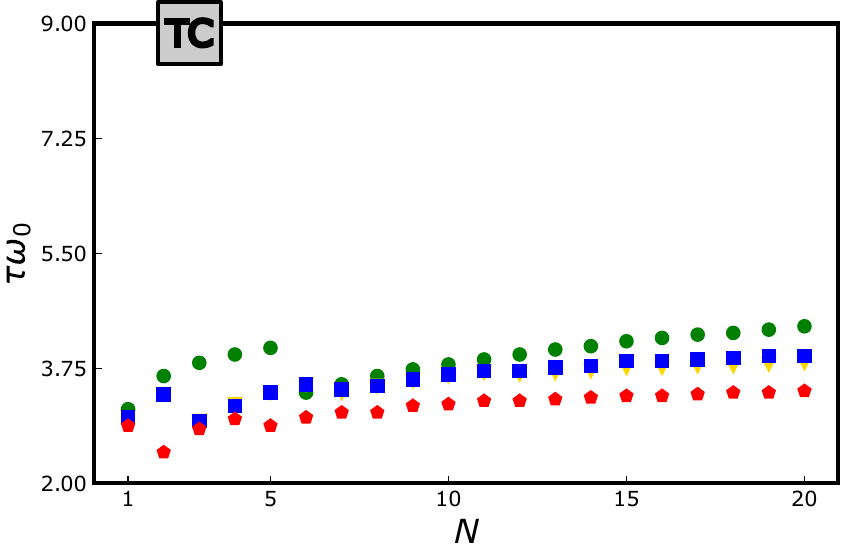}
    \quad
    (k)\includegraphics[width=0.205\linewidth]{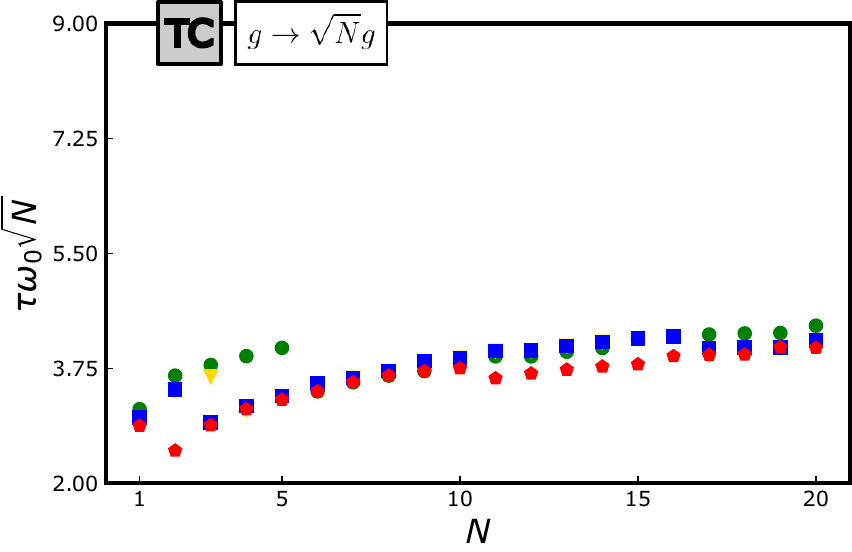}
    \\
    (l)\includegraphics[width=0.205\linewidth]{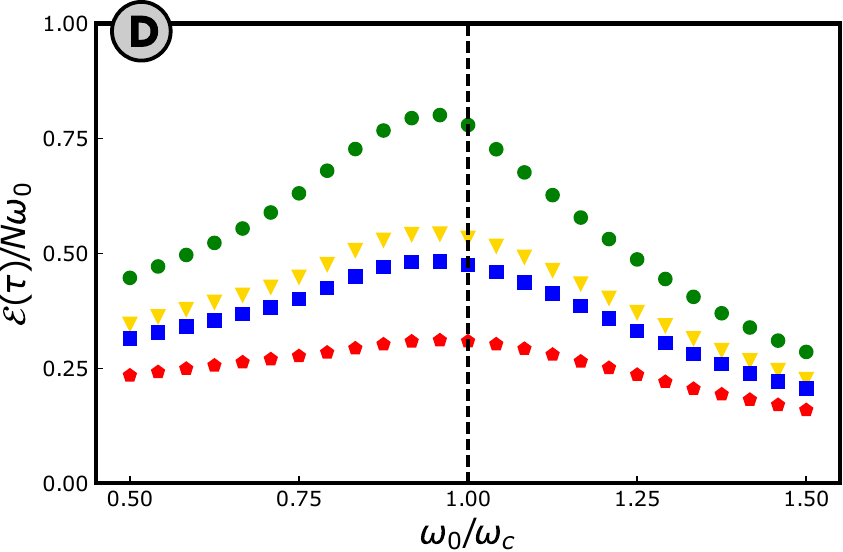}
    \quad
    (m)\includegraphics[width=0.205\linewidth]{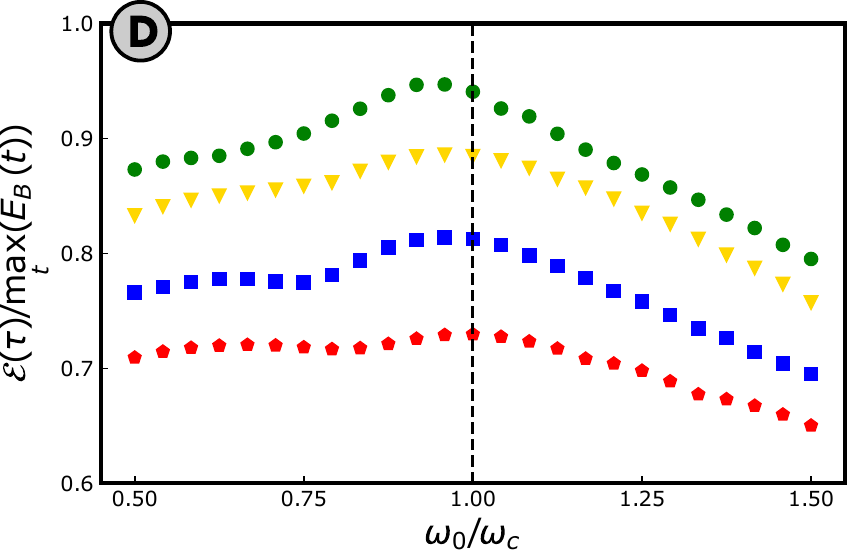}
    \quad
    (n)\includegraphics[width=0.205\linewidth]{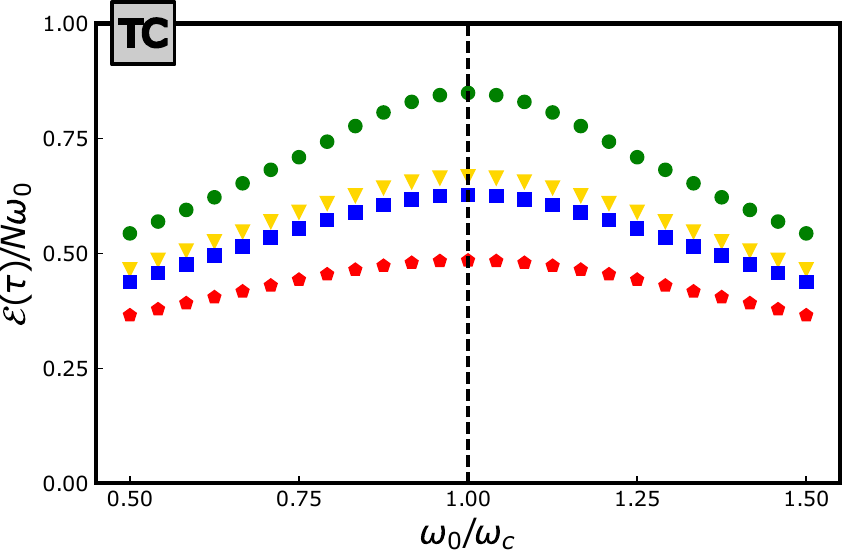}
    \quad
    (o)\includegraphics[width=0.205\linewidth]{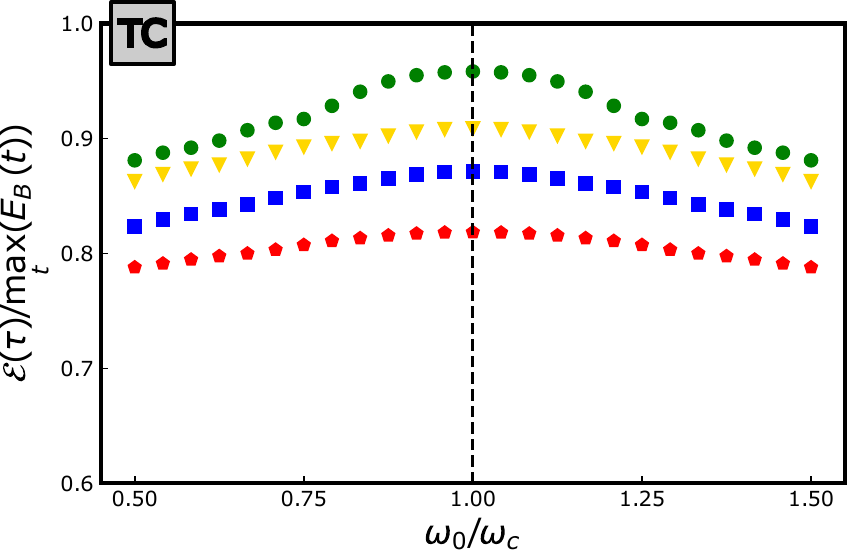}
    \\
     (p)\includegraphics[width=0.205\linewidth]{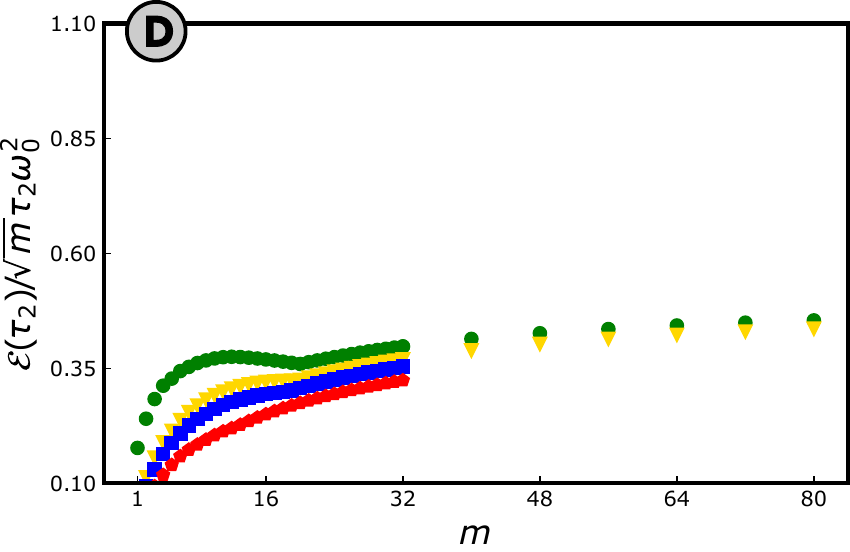}
    \quad
    (q)\includegraphics[width=0.205\linewidth]{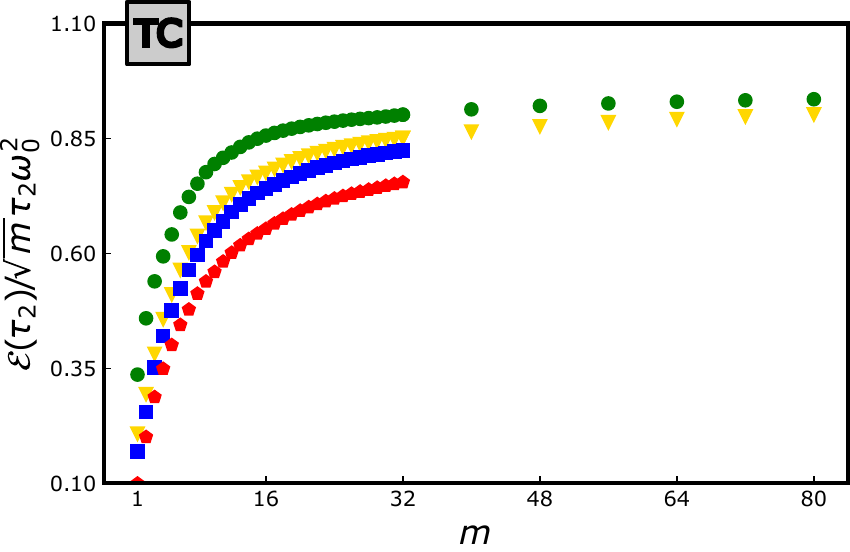}
    \caption{\label{fig:fock} 
        Plots as in Fig.~\ref{fig:extr_energy} (panels from (a) to (d)), \ref{fig:dEon-off} (panels from (e) to (g)), \ref{fig:times} (panels from (h) to (k)), \ref{fig:vsW} (figures from (l) to (o)), \ref{fig:vsM} (panels (p) and (q)) but with charger initially in a Fock state whose mean energy corresponds to the mean energy of the coherent state used in the main text. 
        }
\end{figure*}

All results for the Fock case are displayed in Fig.~\ref{fig:fock}.
Here a commented list of the panels of such figure.
\begin{itemize}
    \item The extractable energy has the same scaling laws and comparable values for both the Dicke (Fig.~\ref{fig:fock} panel (a)) and Tavis-Cummings (Fig.~\ref{fig:fock} panel (c)) interactions.
    \item Instead, the locked energy is higher for both the Dicke (Fig.~\ref{fig:fock} panel (b)) and Tavis-Cummings (Fig.~\ref{fig:fock} panel (d)) interactions.
    Note also that in this case the scaling law of such locked energy resemble the upper bound of Eq.~\eqref{eq:ergBound} found by us.
    This result, combined with the previous one on the ergotropy, suggests that the Fock charging achieve higher mean energies values in the battery, but such extra energy isn't available for extraction.
    \item The quenching mechanism appear to be slightly cheaper in absolute values but more unpredictable in the scaling while charging through a Dicke interaction (Fig.~\ref{fig:fock} panels (f), (g) respectively for $g/\sqrt{N}$, $g$).
    As expected, nothing changes for the Tavis-Cummings interaction (Fig.~\ref{fig:fock} panel (e)).
    \item The charging times are overall the same for the Tavis-Cummings interaction (figures \ref{fig:fock} panels (j), (k)) and a little higher (but always approximately with the same scaling law) for the Dicke interaction (Fig.~\ref{fig:fock} panels (h), (i)) both with the standard normalization $g/\sqrt{N}$ (Fig.~\ref{fig:fock} panels (h), (j)) and bare normalization $g$ (Fig.~\ref{fig:fock} panels (i), (k)).
    \item When we decouple the frequencies of the charger and the battery as expected nothing changes for the Tavis-Cummings interaction (Fig.~\ref{fig:fock} panels (n), (o)).
    Unexpectedly, when charging through the Dicke interaction (Fig.~\ref{fig:fock} panels (l), (m)) we find a result which is opposite to the one found for the coherent charging: now in order to charge faster it is more convenient to exploit more energetic photons than on resonance.
    \item When we increase the amount of energy in the charger, if such energy is initially in a Fock state we recover the same scaling laws found in the main body, both for the Dicke interaction (Fig.~\ref{fig:fock} panel (p)) and Tavis-Cummings interaction (Fig.~\ref{fig:fock} panel (q)).
\end{itemize}

\subsection{Other results}

\begin{figure*}
    (a)\includegraphics[width=0.205\linewidth]{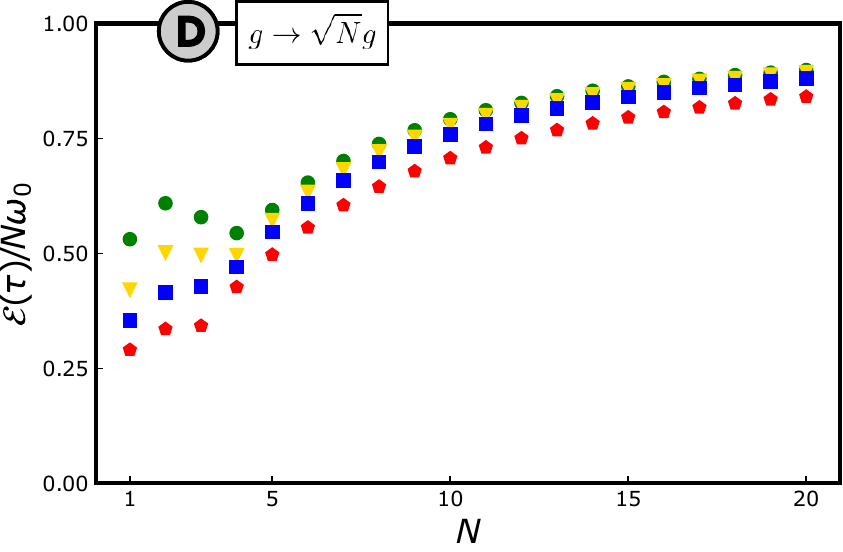}
    \quad
    (b)\includegraphics[width=0.205\linewidth]{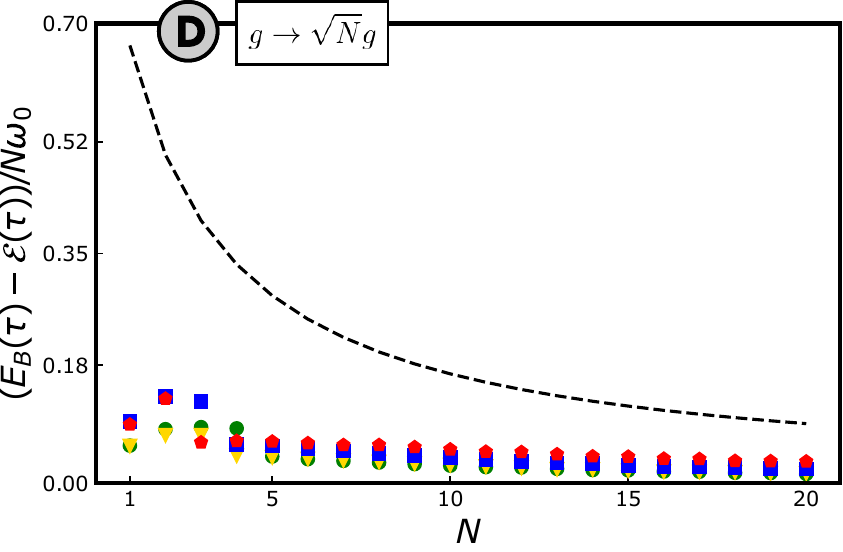}
    \quad
    (c)\includegraphics[width=0.205\linewidth]{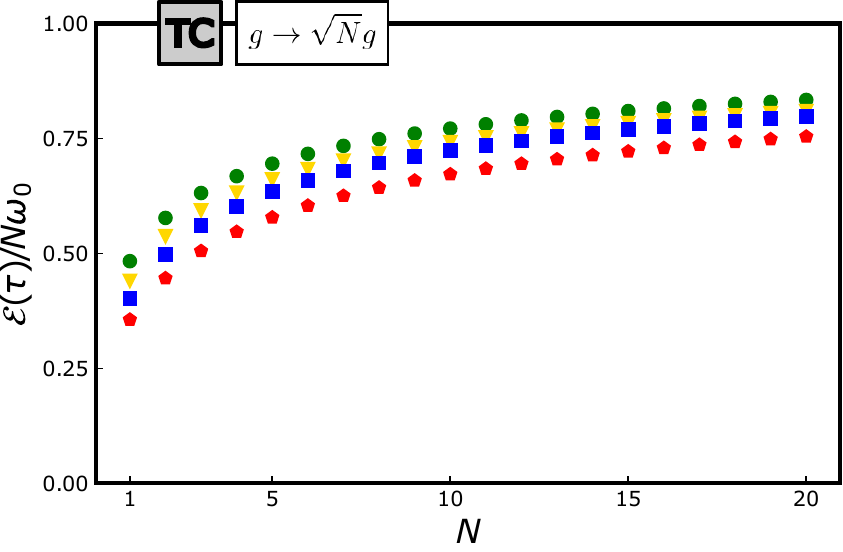}
    \quad
    (d)\includegraphics[width=0.205\linewidth]{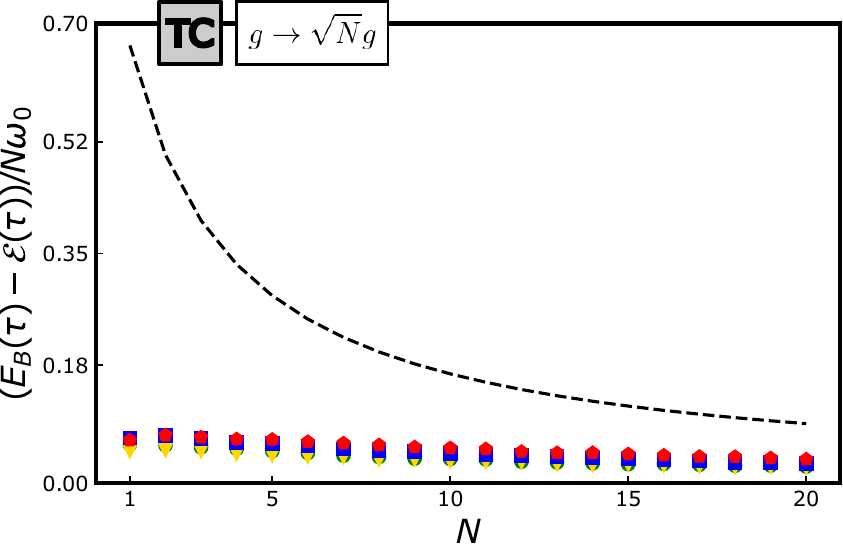}
    \\
    (e)\includegraphics[width=0.205\linewidth]{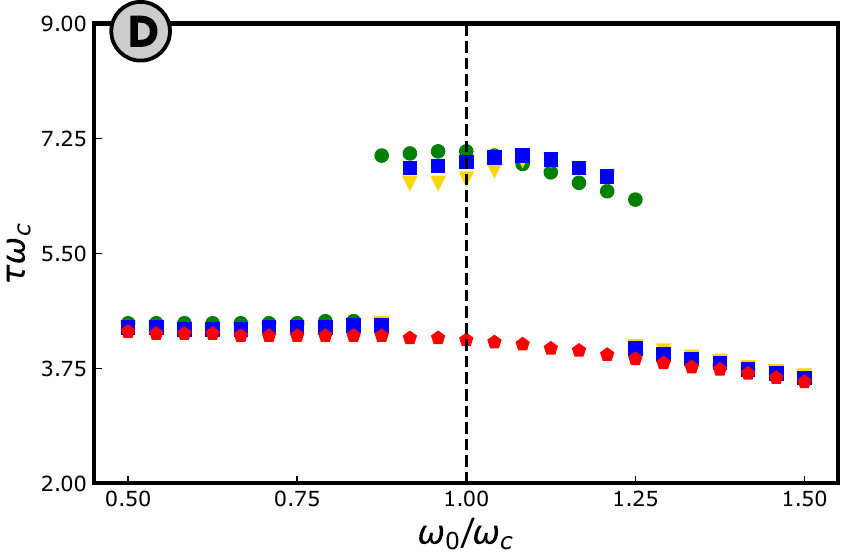}
    \quad
    (f)\includegraphics[width=0.205\linewidth]{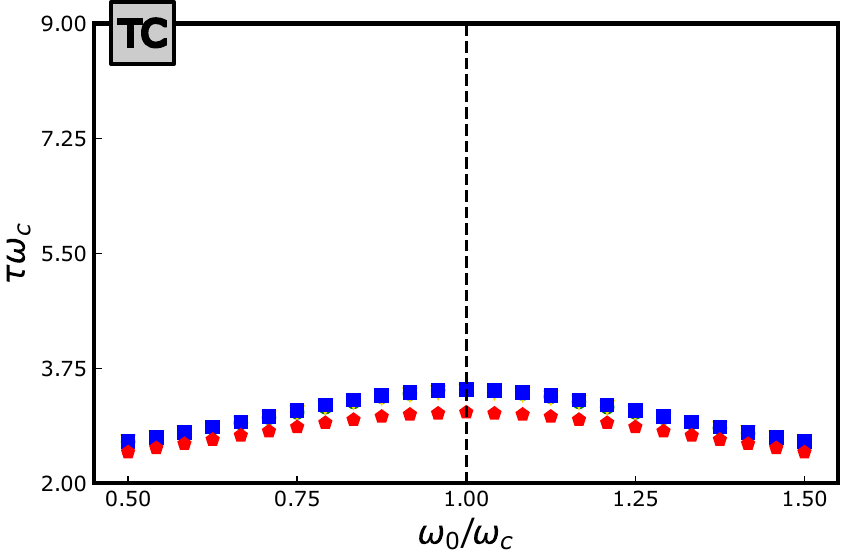}
    \\
    (g)\includegraphics[width=0.205\linewidth]{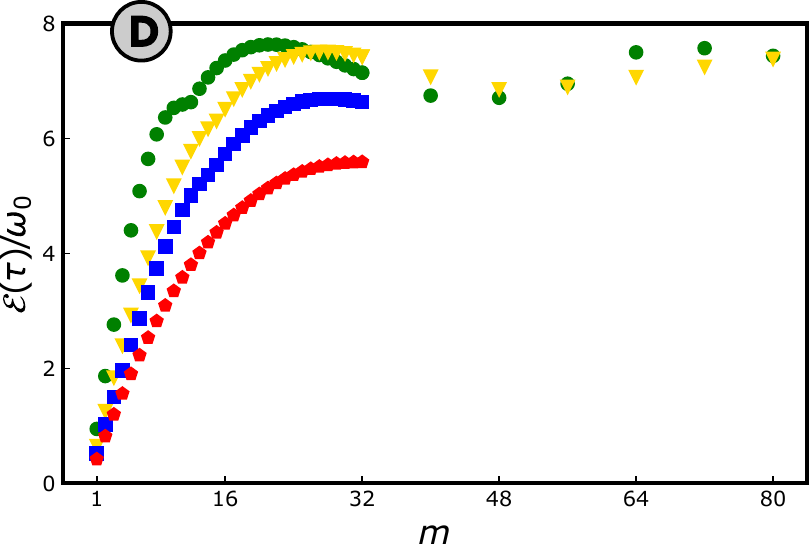}
    \quad
    (h)\includegraphics[width=0.205\linewidth]{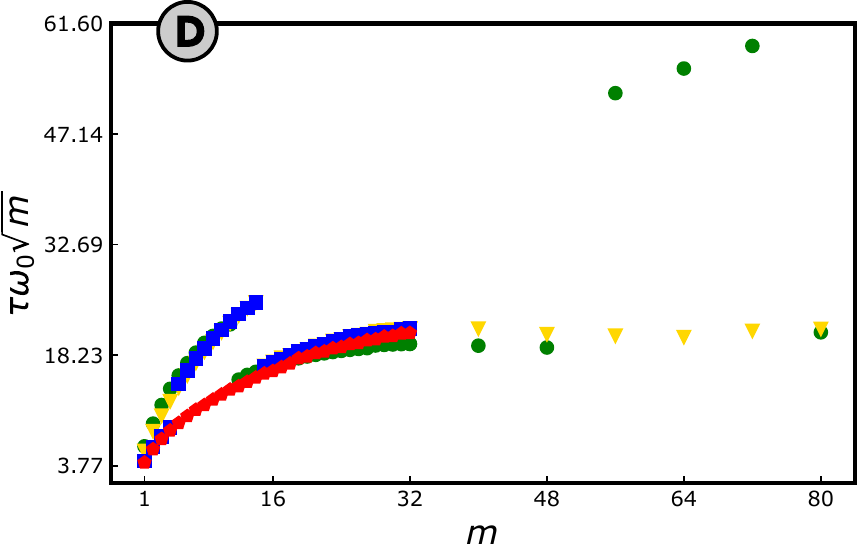}
    \quad
    (i)\includegraphics[width=0.205\linewidth]{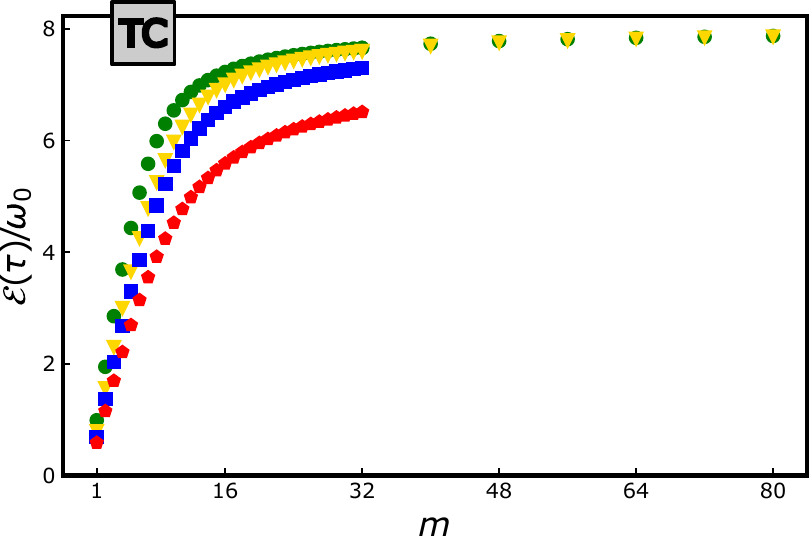}
    \quad
    (j)\includegraphics[width=0.205\linewidth]{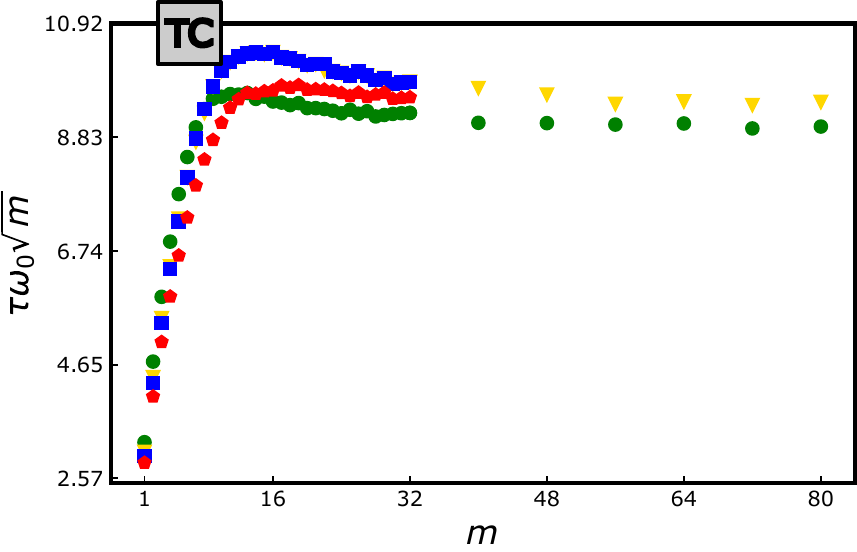}
    \caption{\label{fig:minorres}
        Maximum ergotropy (figures (a), (c)) and locked energy (figures (b), (d)) when charging through a Dicke interaction (figures (a), (b)) and through a Tavis-Cummings interaction (figures (c), (d)), all parameters as in Fig.~\ref{fig:extr_energy} but with the bare interaction normalization $g$.
        Charging time against detuning ratio $\omega_0/\omega_c$, for the Dicke interaction (figures (e)) and Tavis-Cummings interaction (figures (f)), all parameters as in Fig.~\ref{fig:vsW}.
        Maximum ergotropy (figures (g), (i)) and charging time $\tau$ at maximum ergotropy (figures (h), (j)) against the mean energy $m\omega_c$ of the charger, all parameters as in Fig.~\ref{fig:vsM}.
    }
\end{figure*}

Finally, in Fig.~\ref{fig:minorres} we show some minor results mentioned but not displayed in the main body of the article.
We show the results only for the scenario where we charge through the coherent charging, since they display no notable differences in the Fock charging scenario.
If you are interested in the specific values when charging through a Fock state, the figures are available on the GitHub repository \href{https://github.com/TheQuantumJoker/ndqb}{https://github.com/TheQuantumJoker/ndqb}.
Here a commented list of such figure.
\begin{itemize}
    \item In Fig.~\ref{fig:minorres} panels (a) to (d) we can observe the same plot of maximum ergotropy and locked energy as in Fig.~\ref{fig:extr_energy} but with the bare coupling $g$.
    As one can observe with this normalization we still have an approximately extensive amount of extractable energy in the charger and negligible amounts of locked energy.
    The effective values of the energies are higher for the Dicke interaction with this normalization, suggesting that increasing the coupling of the interaction one can increase the amount effectively charged. 
    \item Figure~\ref{fig:minorres} panels (e) and (f) shows the charging time when detuning the frequencies of the battery $\omega_0$ and charger $\omega_c$.
    As showed, detuning the two frequencies results in shorter charging times.
    \item Figure~\ref{fig:minorres} panels (g) and (i) shows the maximum ergotropy when increasing the initial energy of the charger, as in Fig.~\ref{fig:vsM}.
    We observe that when charging through the Dicke interaction (panel (g)) increasing the charger energy do not always lead to more energy charged in the battery, while it is indeed the case when charging through the Tavis-Cummings interaction (panel (i)).
    Figure~\ref{fig:minorres} panels (h) and (j) shows the charging time in the same setup.
    As one can see the law $\tau(m)\propto1/\sqrt{m}$ approximately hold.
\end{itemize}

\end{document}